%
\input lanlmac.tex
\input amssym.def
\input amssym.tex
\overfullrule=0mm
\hfuzz 10pt
%
\def\modif#1{} 

\input epsf.tex
\newcount\figno
\figno=0
\def\fig#1#2#3{
\par\begingroup\parindent=0pt\leftskip=1cm\rightskip=1cm\parindent=0pt
\baselineskip=11pt
\global\advance\figno by 1
\midinsert
\epsfxsize=#3
\centerline{\epsfbox{#2}}
\vskip 10pt
{\bf Fig. \the\figno:} #1\par
\endinsert\endgroup\par
}
\def\hfig#1#2#3{
\par\begingroup
\global\advance\figno by 1
\midinsert
\setbox1=\hbox{\epsfxsize=#3\epsfbox{#2}}%
\vbox to 0pt{%
\vbox to \ht1{\vfill%
\parindent=0pt\leftskip=#3\advance\leftskip by 2cm%
\rightskip=0cm\parindent=0pt%
\baselineskip=11pt%
{\bf Fig. \the\figno:} #1%
\vfill}\vss}%
\noindent\hskip1cm\box1%
\endinsert\endgroup\par
}
\def\figlabel#1{\xdef#1{\the\figno}}
\def\encadremath#1{\vbox{\hrule\hbox{\vrule\kern8pt\vbox{\kern8pt
\hbox{$\displaystyle #1$}\kern8pt}
\kern8pt\vrule}\hrule}}
%
%
%
%
\def\frac#1#2{{\scriptstyle{#1 \over #2}}}

\def\der#1{{\partial \over \partial #1}}

%
%

\def\CG{{\cal G}}              
              
\def\CM{{\cal M}}              \def\CO{{\cal O}}

\def\({ \left( }\def\[{ \left[ }
\def\){ \right) }\def\]{ \right] }
%


\def\IR{\relax{\rm I\kern-.18em R}}
\font\cmss=cmss10 \font\cmsss=cmss10 at 7pt
\def\IZ{\relax\ifmmode\mathchoice
{\hbox{\cmss Z\kern-.4em Z}}{\hbox{\cmss Z\kern-.4em Z}}
{\lower.9pt\hbox{\cmsss Z\kern-.4em Z}}
{\lower1.2pt\hbox{\cmsss Z\kern-.4em Z}}\else{\cmss Z\kern-.4em Z}\fi}
\def\inbar{\,\vrule height1.5ex width.4pt depth0pt}
\def\IB{\relax{\rm I\kern-.18em B}}
\def\IC{\relax\hbox{$\inbar\kern-.3em{\rm C}$}}
\def\ID{\relax{\rm I\kern-.18em D}}
\def\IE{\relax{\rm I\kern-.18em E}}
\def\IF{\relax{\rm I\kern-.18em F}}
\def\IG{\relax\hbox{$\inbar\kern-.3em{\rm G}$}}
\def\IH{\relax{\rm I\kern-.18em H}}
\def\II{\relax{\rm I\kern-.18em I}}
\def\IK{\relax{\rm I\kern-.18em K}}
\def\IL{\relax{\rm I\kern-.18em L}}
\def\IM{\relax{\rm I\kern-.18em M}}
\def\IN{\relax{\rm I\kern-.18em N}}
\def\IO{\relax\hbox{$\inbar\kern-.3em{\rm O}$}}
\def\IP{\relax{\rm I\kern-.18em P}}
\def\IQ{\relax\hbox{$\inbar\kern-.3em{\rm Q}$}}
\def\IGa{\relax\hbox{${\rm I}\kern-.18em\Gamma$}}
\def\IPi{\relax\hbox{${\rm I}\kern-.18em\Pi$}}
\def\ITh{\relax\hbox{$\inbar\kern-.3em\Theta$}}
\def\IOm{\relax\hbox{$\inbar\kern-3.00pt\Omega$}}


\def\oh{{1\over 2}}

\def\Ga{\alpha}\def\Gb{\beta}
\def\Gd{\delta}\def\GD{\Delta}\def\Ge{\epsilon}

\def\Gl{\lambda}

\def\Go{\omega}

\def\diag{{\rm diag \,}}

\def\nind{\noindent}

\def\hepth#1{{\tt hep-th/#1}}\def\mathph#1{{\tt math-ph/#1}}
\def\condmat#1{{\tt cond-mat/#1}}

\def\\#1 {{\tt\char'134#1} }
\def\M{{\widetilde M}}
\catcode`\@=11
\def\Eqalign#1{\null\,\vcenter{\openup\jot\m@th\ialign{
\strut\hfil$\displaystyle{##}$&$\displaystyle{{}##}$\hfil
&&\qquad\strut\hfil$\displaystyle{##}$&$\displaystyle{{}##}$
\hfil\crcr#1\crcr}}\,}   \catcode`\@=12
\def\encadre#1{\vbox{\hrule\hbox{\vrule\kern8pt\vbox{\kern8pt#1\kern8pt}
\kern8pt\vrule}\hrule}}
\def\encadremath#1{\vbox{\hrule\hbox{\vrule\kern8pt\vbox{\kern8pt
\hbox{$\displaystyle #1$}\kern8pt}
\kern8pt\vrule}\hrule}}

\def\l{{\ell}}
\def\He{Heckman} 
\def\GDa{\Delta(a)}  
\def\GDb{\Delta(b)}  
\def\GDal{\Delta_\Gl(a)}  
\def\GDbl{\Delta_\Gl(b)}  
\def\omit#1{}
\def\tr{{\rm tr }} %
\def\Ad{A^\dagger}\def\Bd{B^\dagger}
\def\Ud{U^\dagger}\def\Xd{X^\dagger}
\def\s{\theta}  
\def\t{\bar\theta}  

\def\vertexarr{\epsfxsize=10mm\hbox{\raise -4mm\hbox{\epsfbox{link05.eps}}}}
\def\propagarr{\epsfxsize=14mm\hbox{\raise -1mm\hbox{\epsfbox{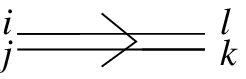}}}}
\def\der{\partial}
\def\d{{\rm d}}
\def\D{{\rm D}}
\def\e#1{{\rm e}^{#1}}
\def\E#1{{\rm e}^{\textstyle #1}}

\def\bra#1{\big< #1 \big|\,}
\def\ket#1{\,\big| #1 \big>}
\def\braket#1#2{\big< #1 \big| #2 \big>}
%
%
%
%
\lref\HC{Harish-Chandra, 
{\sl Differential operators on a semi-simple Lie algebra},
{\it Amer. J. Math.} {\bf 79} (1957) 87--120.}
\lref\We{H. Weyl, {\sl The classical groups}, Princeton University
Press, 1948.}
\lref\IZ{C. Itzykson and J.-B. Zuber, 
{\sl The planar approximation II},
{\it J. Math. Phys.}
{\bf 21} (1980) 411--421.}
\lref\AI{D. Altschuler and C. Itzykson, {\it Ann. IHP} {\bf 54} (1991) 1--8.}
\lref\MSt{M. Stone, Nucl. Phys. {\bf B314} (1989) 557--586. }
\lref\DH{J.J. Duistermaat and  G.J. \He, 
{\sl On the variation of the cohomology of the symplectic form of
the reduced phase space},
{\it Invent. Math.}
{\bf 69} (1982) 259-268,  {\sl Addendum}, {\it ibid} {\bf 72} (1983) 153--158.}
\lref\Mor{A. Morozov, 
{\sl Pair Correlator in the Itzykson-Zuber Integral},
{\it Mod. Phys. Lett.} {\bf A7} (1992) 3503--3508, \hepth{9209074}.}
\lref\Maty{ A. Matytsin, 
{ \sl On the Large N Limit of the Itzykson-Zuber Integral},
{\it Nucl. Phys.} {\bf B411} (1994) 805--820, \hepth{9306077}.}
\lref\Shat{S. Shatashvili, 
{\sl  Correlation Functions in The Itzykson-Zuber Model}, 
{\it Commun. Math. Phys.} {\bf 154} (1993) 421--432,  \hepth{9209083}.}
\lref\susy{
J. Alfaro, R. Medina and L. F. Urrutia,
{\sl Itzykson-Zuber Integral for $U(m|n)$},
{\it J. Math. Phys.} {\bf 36} (1995) 3085--3093, \hepth{9412012}\semi
T. Guhr, {\it Commun. Math. Phys.} {\bf 176} (1996) 555--586.}
\lref\GW{T. Guhr and T. Wettig, 
{\sl An Itzykson-Zuber-like Integral and Diffusion for Complex 
 Ordinary and Supermatrices}, 
{\it J. Math. Phys.} {\bf  37} (1996) 6395--6413,
\hepth{9605110}.}
\lref\Fyo{Y.V. Fyodorov and E. Strahov,
{\sl Characteristic polynomials of random Hermitian matrices and 
Duistermaat-Heckman localisation on non-compact Kaehler manifolds},
{\it Nucl. Phys.} {\bf B630} (2002) 453--491, \mathph{0201045}.}
\lref\JSV{A.D. Jackson, M.K. \c Sener, J.J.M. Verbaarschot,
{\sl Finite volume partition functions and Itzykson-Zuber integrals},
{\it Phys. Lett.} {\bf B387} (1996) 355--360, \hepth{9605183}.}

\lref\Ba{A.B. Balantekin, 
{\sl Character Expansions, Itzykson-Zuber Integrals, 
and the QCD Partition Function}, 
{\it Phys. Rev.} {\bf D62} (2000) 085017,  \hepth{0007161}.} 
\lref\fourtout{
 T. Akuzawa and  M. Wadati,  
{\sl Effective QCD Partition Function in Sectors with Non-Zero 
Topological Charge and Itzykson-Zuber Type Integral}, 
J. Phys. Soc. Jap. 67 (1998) 2151--2154, hep-th/9804049
}
\lref\BaCa{A.B. Balantekin and P. Cassak, \hepth {0108130}.} 
\lref\Sz{R. J. Szabo,
{\sl Equivariant Localization of Path Integrals},
\hepth{9608068}\semi
Y. Karshon, {\sl Lecture notes on Group Actions on Manifolds}
(1996--97), \hfill\break
{\tt http://www.ma.huji.ac.il/$\sim$karshon/teaching/1996-97/actions/lecture-notes.html}.}
\lref\Jacobid{R. Vein and P. Dale, 
{\sl Determinants and their applications in mathematical physics},   
Springer 1999\semi
D. M. Bressoud, {\sl Proofs and Confirmations}, Cambridge Univ. Pr. 1999.
}
\lref\Hua{L.K. Hua, {\sl 
Harmonic analysis of functions of several complex variables 
in the classical domains}, AMS, Providence 1963.}
%
\lref\Ivan{I.K. Kostov, 
{\sl String Equation for String Theory on a Circle},
{\it Nucl. Phys.} {\bf B624} (2002) 146--162,
\hepth{0107247}.}
\lref\PZJb{P. Zinn-Justin,
{\sl Universality of correlation functions of hermitian random
matrices in an external field},
{\it Commun. Math. Phys.} {\bf 194}, 631--650 (1998).}
\lref\PZJ{P. Zinn-Justin, 
{\sl HCIZ integral and 2D Toda lattice hierarchy}, 
{\it Nucl. Phys.} {\bf B634} (2002) 417--432,
\mathph{0202045}.}
\lref\KM{V.A.~Kazakov and A.A.~Migdal,
{\it Nucl. Phys.} {\bf B397} (1993) 214--238.}
\lref\UT{K.~Ueno and K.~Takasaki, {\it Adv. Stud. Pure Math.} 4,
editors H.~Morikawa and K.~Okamoto (1984), 1.}
\lref\BC{B. Collins, {\sl Moments and cumulants of polynomial
random variables on unitary groups, the Itzykson-Zuber integral 
and free probability}, \mathph{0205010}.}
\lref\BMS{M. Bousquet-M\'elou and G. Schaeffer, 
{\sl Enumeration of planar constellations}, 
{\it Adv. in Appl. Math.} 24(4) (2000)
337--368\semi 
G. Schaeffer, {\sl Conjugaison d'arbres et cartes combinatoires
al\'eatoires}, Th\`ese de doctorat,
{\tt http://dept-info.labri.u-bordeaux.fr/$\sim$schaeffe/cv/bibli/These.html}}
\lref\OBZ{H. O'Brien and J.-B. Zuber, {\it Phys. Lett} {\bf 144B}
(1984) 407--408;  {\it Nucl. Phys.} {\bf B253} (1985) 621--634.}
\lref\GMGW{T.~Guhr, A.~Mueller-Groeling
and H.A.~Weidenmueller,
{\sl Random Matrix Theories in Quantum Physics: Common Concepts},
{\it Phys. Rep.} 299 (1998), 189--425, \condmat{9707301}.}
\lref\PDF{P.~Di~Francesco, {\sl Rectangular Matrix Models and Combinatorics of Colored Graphs}, \condmat{0208037}.}
\lref\KSW{V.A.~Kazakov, M.~Staudacher and T.~Wynter,
{\it Commun. Math. Phys.} {\bf 177} (1996), 451--468; {\bf 179}
(1996), 235--256;
{\it Nucl. Phys.} {\bf B471} (1996), 309--333.}
%
%
\lref\BIZ{D. Bessis, C. Itzykson and J.-B. Zuber, 
{\sl Quantum Field Theory Techniques in Graphical Enumeration},
{\it Adv. Appl. Math.} {\bf 1} (1980) 109--157.}
\lref\BIPZ{E. Br\'ezin, C. Itzykson, G. Parisi and J.-B. Zuber, 
{\sl Planar Diagrams}, 
{\it Commun. Math. Phys.} {\bf 59} (1978) 35--51.}
\lref\Spe{R. Speicher, {\sl Math. Ann.} {\bf 298} (1994) 611-628.}
\lref\DFGZJ{P. Di Francesco, P. Ginsparg and J. Zinn-Justin, 
{\sl 2D Gravity and Random Matrices, }
{\it Phys. Rep.} {\bf 254} (1995) 1--133.}
\lref\tH{G. 't Hooft, 
{\sl A Planar Diagram Theory for Strong Interactions}, 
{\it Nucl. Phys.} {\bf B 72} (1974) 461--473.}
\lref\KP{V.A.~Kazakov and P.~Zinn-Justin, 
{\sl Two-Matrix Model with $ABAB$ Interaction}, 
{\it Nucl. Phys.} {\bf B 546} (1999) 647--668.}
\lref\Tutte{W.T.~Tutte, {\sl A Census of Planar Maps}, 
{\it Can. J. Math.} {\bf 15} (1963) 249--271.}
\lref\ZJZ{P.~Zinn-Justin and J.-B. Zuber, 
{\it J. Knot Theor. Ramif.} 9 (2000), 1127, \mathph{0002020}.}
\lref\Zv{A.~Zvonkin, 
{\sl Matrix Integrals and Map Enumeration: An Accessible Introduction},
{\it Math. Comp. Modelling} {\bf 26} (1997) 281--304.}
\lref\Ja{A.T.~James, {\it Ann. Math. Statist.} 35 (1964), 475--501.}
\lref\GZ{A.~Guionnet, O.~Zeitouni, {\sl Large deviations asymptotics
for spherical integrals}, to appear in {\it J. Func. Anal.},
{\tt http://www.umpa.ens-lyon.fr/$\sim$aguionne/}.
}
\Title{
}
{{\vbox {
\vskip-10mm
\centerline{
On some integrals over the $U(N)$ unitary group} 
\centerline{
and their large $N$ limit}
}}}
\medskip
\centerline{P. Zinn-Justin}\medskip
\centerline{\it Laboratoire de Physique Th\'eorique et Mod\`eles Statistiques}
\centerline{\it Universit\'e Paris-Sud, B\^atiment 100}
\centerline{\it F-91405 Orsay Cedex, France}
\bigskip
\centerline{\sl and}
\medskip
\centerline{J.-B. Zuber}\medskip
\centerline{\it Service de Physique Th\'eorique de Saclay}
\centerline{\it CEA/DSM/SPhT, Unit\'e de recherche associ\'ee au CNRS}
\centerline{\it CEA-Saclay, F-91191 Gif sur Yvette Cedex, France}

\vskip .2in

\noindent 
The integral over the $U(N)$ unitary group 
$I=\int \D U \,\exp\Tr A U B U^\dagger$
is reexamined. Various approaches and extensions are first reviewed.
The second half of the paper deals with more recent developments: 
relation with integrable Toda lattice hierarchy, diagrammatic 
expansion and combinatorics, and what they teach us on the large 
$N$ limit of $\log I$. 
\bigskip

\bigskip
\vskip25mm

\noindent{Submitted to J. Phys. A, 
Special Issue on Random Matrix Theory }
\Date{09/02} 
%
\vfill\eject
\omit{\input sec1
\input sec2
\input sec3
\input sec4
\input sec5
\input sec6}
\newsec{Introduction and notations}
\subsec{Aim and plan of the paper}
\noindent 
\def\hepph#1{{\tt hep-ph/#1}}
\lref\VW{J.J.M. Verbaarschot and T. Wettig, 
{\sl Random Matrix Theory and Chiral Symmetry in QCD }
{\it Ann. Rev. Nucl. Part. Sci.} {\bf 50} (2000) 343-410,
\hepph{0003017}.}
\lref\DFmsri{P. Di Francesco, 
{\sl Random Matrices and Their Applications},
{\it MSRI Publications} {\bf 40} (2001)
111-170 (2001) 
Bleher P.M., Its A.R., eds. (Cambridge University Press), 
\mathph{9911002}}
\lref\Morb{A. Morozov, 
{\sl Integrability and Matrix Models}, 
{\it Phys. Usp.} {\bf 37} (1994) 1-55,
\hepth{9303139}.}
\lref\AvM{
 M. Adler and P. van Moerbeke, 
{\sl String-Orthogonal Polynomials, String equations, and 2-Toda symmetries},
{\it Commun. Pure Appl. Math.} {\bf 50} (1997) 241-290, \hepth{9706182}.}
\lref\OS{
A. Yu. Orlov and D. M. Scherbin, {\sl Fermionic representation
for basic hypergeometric functions related to Schur polynomials},
\mathph{0001001}\semi
A. Yu. Orlov, {\sl Tau Functions and Matrix Integrals},
\mathph{0210012}.}
\lref\VDN{D.V. Voiculescu, K.J. Dykema and A. Nica,
{\sl Free random variables}, 
CRM monograph series (AMS, Providence RI, 1992).}
Techniques of integration over large matrices are important 
in several contexts of physics -- from QCD \VW\ to quantum gravity
\DFGZJ, from disordered systems to
mesoscopic physics \GMGW -- 
and of mathematics -- enumerative combinatorics \DFmsri,
integrable systems \refs{\Morb,\AvM,\OS}, free probability theory \VDN, 
statistics \Ja, etc. 
In most applications, a central r\^ole is played by
the statistics of the eigenvalues of the random 
matrices -- spectrum  and correlations -- or more generally 
of their invariants $\Tr A^p$. This is justified once ``angular''
variables have been integrated over, like those appearing in 
$\Tr AB$. 

It is in this general context that the integral 
$$I=\int_{U(N)} \D U \exp \Tr A U B \Ud\ ,$$
$A$ and $B$ hermitian, 
was studied more than twenty years ago, and an exact expression  
was derived \IZ. Soon after, it was realised that this result had been
obtained long before by Harish-Chandra \HC\
as a corollary of a more general problem. 
The purpose of this paper is to return to this
integral (sometimes called the Harish-Chandra--Itzykson--Zuber
integral in the physics literature),
to review the known facts and to present some of its known
extensions, before turning to what remains a challenge: to find a 
good and systematic description of the large $N$ limit of its 
{\it logarithm}. We shall report on some recent progress made on the latter
issue. 

Our paper is organised as follows. 
In the rest of sec.~1, we introduce  notations
and some basic results. 
The derivation of the expression of $I$ is then reviewed in sec.~2,  
using various methods:  heat equation,  
character expansion and Duistermaat-\He\ theorem. 
Sec.~3 discusses briefly extensions in various directions, in
particular the case of rectangular matrices.
Connection with integrable hierarchies is then presented in 
sec.~4, with special attention devoted to its dispersionless limit
and what can be learnt from it. The resulting expressions for the
large $N$ limit of $\log I$ are then confronted in sec.~5 to those obtained
by (what we believe to be) a novel diagrammatic expansion of 
$\log I$ and by a purely combinatorial analysis \BC.
Finally sec.~6 contains a summary of results and tables. 
 
\subsec{Notations}
\nind Let $A$ and $B$ be two $N\times N$ matrices.
We shall assume that they are Hermitean, even though many properties
that we shall derive do not require it.
The subject of study is the integral
\eqn\hciz{I(A,B;s)=\int_{U(N)} \D U \exp  {N\over s} \Tr AUB\Ud}
where $\D U$ is the Haar measure on the unitary group $U(N)$,
normalised to $\int \D U=1$;
and the large $N$ limit (in a sense defined below)
of its logarithm ${1\over N^2}\log I(A,B;s)$. At the 
possible price of a redefinition of $U$, one may 
always assume that $A$ and $B$ are diagonal, 
$A=\diag(a_i)_{1\le i\le N}$, $B=\diag(b_i)_{1\le i\le N}$.
Clearly, the real parameter $s$ could be scaled away. We find
convenient to keep it as an indicator 
of the homogeneity in the $a$'s and $b$'s.

To any order of their $1/s$ expansions,  both $I(A,B;s)$ and
its logarithm are completely symmetric polynomials in the
eigenvalues $a_i$ and in the $b_i$ independently,
as we shall see in the next section. We want to express them in terms 
of the elementary symmetric functions of the $a$'s and of the $b$'s:
\eqn\symmfn{\s_p:={1\over N} \sum_{i=1}^N a_i^p\qquad 
\t_p:={1\over N}\sum_{i=1}^N b_i^p\ .}
%
For finite $N$, only a finite number of these functions are
independent, but this constraint disappears as $N\to \infty$ .
By large $N$ limit we therefore mean that we consider sequences
of matrices $A$ and $B$ of size $N$ such that the moments of their spectral
distributions $\s_p$ and $\t_p$ converge as $N\to\infty$
(see \GZ\ for some rigorous results on this limit). By abuse of notation
we still denote by $A$ and $B$ such objets, and write
\eqn\freeen{F(A,B;s)=\lim_{N\to\infty} {1\over N^2}\log I(A,B;s)\ .}

We shall use various combinations of the
symmetric functions $\s_p$ and $\t_p$. In particular, 
for  $\Ga\vdash n$, i.e. $\Ga$ a partition of 
$n=\sum_p p\,\Ga_p$,  which we 
also write $\Ga=[1^{\Ga_1}\cdots n^{\Ga_n}]$, we define
\eqn\defi{\tr_\Ga A:= \Big({1\over N}\Tr A\Big)^{\Ga_1}\cdots 
\Big({1\over N} \Tr A^n\Big)^{\Ga_n}=\prod_{p=1}^n \theta_p^{\,\alpha_p} \ .}
We shall also make use of
the characters of the irreducible holomorphic 
representations of the linear group GL$(N)$, 
labelled by Young diagrams $\Gl$ with rows of lengths 
$0\le \Gl_1\le\Gl_2\le \cdots \le \Gl_N$. 
They read 
\eqn\chara{\chi_\Gl= {\GDal\over \GDa}}
in terms of Vandermonde determinants 
\eqn\Vandm{\GDa=\prod_{1\le j<i\le N} (a_i-a_j)=\det(a_i^{j-1})}
and their generalizations
\eqn\Vandmg{\GDal= \det(a_i^{\Gl_j+j-1})\ .}

Frobenius formula \We\ relates these sets of symmetric functions:
if $\Ga\vdash n$
\eqn\frob{N^{\Sigma \Ga_j}\, \tr_{\Ga} A =
\sum_{\Gl\atop |\Gl|=n} \chi_\Gl(A) \hat\chi_\Gl(\Ga)}
where the sum runs over all Young tableaux $\Gl$ with $|\Gl|=n$ boxes, 
and
$\hat\chi_\Gl(\Ga)$ denotes the character of the symmetric group
$\goth{S}_n$, 
for the representation labelled by 
$\Gl$ and for the class labelled by $\Ga$.

\bigskip
The expression of integrals and differential operators over
 hermitian matrices in terms of their eigenvalues involves 
a Jacobian, and problems of normalisation appear. We discuss 
these questions shortly. 

If $M=U A \Ud$, $A$ diagonal, 
$U$ unitary, we have $\d M= U \d A \Ud+ [\d X,M]$, 
where $\d X:= \d U\Ud$ is antihermitian. Then
$\Tr (\d M)^2= \sum_i \d a_i^2 +2\sum_{i<j} |\d X_{ij}|^2 |a_i-a_j|^2$
defines the metric tensor $g_{\Ga\Gb}$ in the coordinates 
$\xi^\Ga=(a_i, X_{ij})$. This determines first the measure
$\D M= \sqrt{\det g} \prod \d\xi^\Ga=
2^{N(N-1)/2} \prod_i \d M_{ii}\,\prod_{i<j}\,\d\Re e M_{ij}\,\d\Im m  M_{ij}=
 2^{N(N-1)/2}\Delta^2(a) 
\prod \d a_i \prod \d X_{ij}$ $=C \Delta^2(a) \prod_i \d a_i\,\D U$. 
The constant $C$ is fixed by computing in two different ways
the integral of a $U(N)$ invariant function of $M$, for example a Gaussian
$$1=\int \D M {\e{-\oh \Tr M^2}\over (2\pi)^{N^2/2}}=
{C\over (2\pi)^{N(N-1)/2}} \int \prod_{i=1}^N {\d a_i\over (2\pi)^{\oh}}
\e{-\oh  a_i^2} \prod_{i<j} (a_i-a_j)^2 
\ .
$$
The latter integral equals $\prod_{p=1}^N p!$ thus
\eqn\consta{C={(2\pi)^{N(N-1)/2}\over \prod_{p=1}^N p!}\ ,\qquad 
\D M={(2\pi)^{N(N-1)/2}\over \prod_{p=1}^N p!} \Delta^2(a)
\prod_{i=1}^N \d a_i\, \D U\ .}
 
From the metric above, one also computes the Laplacian 
\eqn\lapla
{\eqalign{\Delta_M &={1\over \sqrt{g}}
{\partial \over \partial \xi^\Ga} g^{\Ga\Gb}\sqrt{g} 
{\partial \over \partial \xi^\Gb}
= {1\over \prod_{i<j} (a_i-a_j)^2} \sum_k {\partial \over \partial
a_k}  \prod_{i<j} (a_i-a_j)^2 {\partial \over \partial a_k}
+\Delta_X 
\omit{ \cr &= \sum_i {\partial^2 \over \partial a_i^2}
+2 \sum_{i<j} {1\over a_i-a_j}\({\partial \over \partial a_i}-
{\partial \over \partial a_j}\) +\Delta_X }
\cr &= \GDa^{-1} \sum_k  \Big({\partial \over \partial a_k}\Big)^2
\GDa
+\Delta_X \ ,
}}
where the last equality results from the vanishing of
$\sum_k [\partial_{a_k}, [\partial_{a_k},\GDa]]=0$,
as this completely antisymmetric function of the $a$'s is a polynomial
of degree $N(N-1)/2 -2$.

%
\newsec{The exact expression of $I(A,B;s)$}
\noindent
Assume that all the eigenvalues of $A$ and $B$ are distinct.
One finds \refs{\HC,\IZ}
\eqn\resu{I(A,B;s)= 
\bigg( \prod_{p=1}^{N-1}p!\bigg)\,  (N/s)^{-N(N-1)/2} \,
{\det\big(\E{{N\over s} a_i b_j}\big)_{1\le i,j\le N}
\over \GDa\GDb\ }}
Note that both the numerator and the denominator of the r.h.s.\ 
are completely antisymmetric functions of the $a$'s and of the $b$'s
independently, and that the limit where some eigenvalues coalesce 
is well defined.

This expression \resu\ may be obtained by 
several different routes.

\subsec{Heat equation} 
\nind For two hermitian $N\times N$ matrices $M_A$ and $M_B$,  let us consider 
\eqn\green{K(M_A,M_B;s)=  \({N\over 2\pi s }\)^{N^2/2}\,
\exp  -{ N\over 2s} \Tr (M_A-M_B)^2\ .}
$K(M_A,M_B;s)$ satisfies the heat equation 
\eqn\heat{\({\partial\over \partial s}-\oh\Delta_{M_A}\)K(M_A,M_B;s) =0\
,}
where $\GD_M$ is the Laplacian over $M$,
together with the boundary condition that for $t\to 0$, 
$K(M_A,M_B;s)\to \delta(M_A-M_B)$.

The heat kernel 
$K(M_A,M_B;s)$ is invariant under the simultaneous adjoint action on 
$M_A$ and $M_B$ by the same unitary matrix $U$. If we
diagonalise $M_A=U_A A \Ud_A$ and $M_B=U_B B \Ud_B$, 
$K(M_A,M_B;s)= K(A,U B\Ud;s)$ where $U=\Ud_A U_B$. 
Upon integration
\eqnn\integr
$$\eqalignno{
\widetilde K(A,B;s)&:=\int \D U\, K(M_A, U M_B \Ud;s)= \int \D U\, K(A, U B
\Ud;s) 
\cr &=
 \({N\over 2\pi s }\)^{N^2/2}\,   \E{- {N\over 2s} \Tr (A^2+B^2)}
I(A,B;s)
& \integr
}$$
is again a solution of the heat equation \heat, but depends only on 
the eigenvalues $a_i$ of $A$ (and $b_i$ of $B$). Using the explicit form 
of the Laplacian in terms of the eigenvalues $a_i$,  (see \lapla), 
 $\widetilde K$  satisfies
\eqn\heattil{  \({\partial\over \partial s}-\oh \sum_k \({\partial \over 
\partial a_k}\)^2 \) \GDa\, \widetilde K(M_A,M_B;s) =0\ .}
The product $\GDa \GDb \widetilde K(A,B;s)$
 is an antisymmetric function of the $a$'s and of the $b$'s, is a 
solution of the heat equation with the flat Laplacian, 
and satisfies the boundary conditions that for $s\to 0$, 
$C \GD(a) \GD(b) \widetilde K(A,B;s)\to {1\over N!}\sum_{P\in
\goth{S}_N} \Ge_P \prod_i
\Gd(a_i-b_{Pi})$, with $C$ the constant computed in \consta. 

In physical terms it is the Green function of $N$ independent
free fermions, and it is thus given by the Slater 
determinant
\eqn\slater{ 
C  \GD(a) \GD(b) \widetilde K(A,B;s)=
{1\over N!} \({N\over 2\pi s}\)^{N/2}
\, \det\left[ \exp -{N\over 2s} (a_i-b_j)^2\right]\ ,
}
which is consistent with \resu. 

\subsec{Character expansion}
\nind We may expand the exponential in \hciz\ to get
\eqn\powexp{I(A,B;s)=\sum_{n=0}^\infty {(N/s)^n\over n!} \int \D U  
(\Tr AUB\Ud)^n\ .}
Frobenius formula \frob\ evaluated for the partition $[1^n]$
gives for any matrix $X$ of GL$(N)$ 
\eqn\trn{
\Tr^n X=\sum_{\Gl\atop |\Gl|=n} \hat d_\Gl \chi_\Gl(X) \ ,
}
where  $\hat d_\Gl=\hat\chi_\Gl([1^n])$ is the dimension of the
$\Gl$-representation of $\goth{S}_n$. 
Integration over the unitary group then yields
\eqn\orthog{
 \int \D U \chi_\Gl(AUB\Ud)
={\chi_\Gl(A)\chi_\Gl(B)\over\chi_\Gl(I)}={\chi_\Gl(A)\chi_\Gl(B)\over
d_\Gl}
}
and a well known formula \We\ gives 
\eqn\well{
 {\hat d_\Gl\over d_\Gl}= n! \prod_{p=1}^{N}{(p-1)!\over (\Gl_p+p-1)! }\ .
}
Using \chara\ and putting everything together we find
\eqna\togeth
$$\eqalignno{\!\!\!\!\!\!\!\!
\GDa\,\GDb\, I(A,B;s)&=\Big(\prod_{p=1}^{N-1} p!\Big)
\sum_{n=0}^\infty (N/s)^n
\sum_{\Gl\atop |\Gl|=n}
{1\over\prod_p (\Gl_p+p-1)!} \GDal\GDbl& \togeth a\cr
&=\Big(\prod_{p=1}^{N-1} p!\Big) 
\sum_{0\le \l_1<\cdots\l_N}
\prod_{q=1}^N{(N/s)^{\l_q-q+1}\over (\l_q)!} \det(a_i^{\l_j})
\det(b_i^{\l_j}) & \togeth b \cr
}$$ 
Eq. \togeth{a} is interesting in its own sake, 
while an extension of Binet--Cauchy theorem%
%
\foot{Recall \Ba\ that if $f(x)=\sum_{\l\ge 0} f_\l x^\l$, 
$$\eqalign{\sum_{0\le \l_1<\l_2<\cdots<\l_N} 
\!\!\!\!\!\!\! f_{\l_1} \cdots f_{\l_N}
\det a_i^{\l_j}\det b_i^{\l_j}
&={1\over N!} \sum_{\l_i\ge 0}
\sum_{P,P'} \Ge_P\,\Ge_{P'} f_{\l_1} \cdots f_{\l_N}\prod_i
a_{Pi}^{\l_i}b_{P'i}^{\l_i}\cr
&={1\over N!}\sum_{P,P'} \Ge_{P.P'}\prod_{i}\sum_{\l_i}f_{\l_i}
(a_{P.P'i} b_i)^{\l_i} \cr
&=\det f(a_i b_j)
}$$}
 enables one to resum \togeth{b} into
\eqn\resub{
\GDa\,\GDb\, I(A,B;s)=\Big(\prod_{p=1}^{N-1} p!\Big)  \,(N/s)^{-N(N-1)/2}
\, \det \big(\e{ {N\over s} a_i b_j}\big)
}
which is precisely \resu.

\subsec{Duistermaat-\He\ theorem}
\nind
Let us compute the integral \hciz\ 
by the stationary phase method. 
The stationary points $U_0$ of the ``action'' $\Tr AUB\Ud$
satisfy $ \Tr \,\delta U U_0^\dagger\, [U_0 B U_0^\dagger,A ]=0$
for arbitrary antihermitian $\delta U U_0^\dagger$, hence 
$ [U_0 B U_0^\dagger,A]=0\ .$  For diagonal matrices
$A$ and $B$ with distinct eigenvalues, this implies 
that $U_0 B U_0^\dagger$ is diagonal 
and therefore that the saddle point $U_0$ are permutation matrices. 

Gaussian fluctuations around the stationary point $U_0=P$
may be computed by writing $U= \e{X} P$, $X$ antihermitian, and 
by integrating over $X$ after expanding the action to second order. 
Summing over all stationary points
thus gives the ``one-loop approximation'' to the integral \hciz: 
\eqnn\oneloop
$$\eqalignno{
I(A,B;s)
&=C'\sum_{P\in \goth{S}_N}
\e{ {N\over s} \sum_{i=1}^N a_i b_{Pi}} 
\int \prod_{i<j} d^2 X_{ij}\,
\e{ -{N\over s} \sum_{i<j} |X_{ij}|^2 (a_i-a_j)(b_{Pi}-b_{Pj})}\cr
&=C' \sum_{P\in \goth{S}_N}
\e{ {N\over s} \sum_{i=1}^N a_i b_{Pi}} 
{\({\pi s\over  N}\)^{N(N-1)/2} \over 
\prod_{i<j}  (a_i-a_j)(b_{Pi}-b_{Pj})} \cr
&=C' \({s \pi \over N}\)^{N(N-1)/2} {1\over \GDa \GDb}
 \sum_{P\in \goth{S}_N} \Ge_P\, 
\e{ {N\over s} \sum_{i=1}^N a_i b_{Pi}} \cr
&= C' \({s \pi \over N}\)^{N(N-1)/2} 
{\det \Big(\E{{N\over s} a_i\,b_j} 
\Big)\over \GDa \GDb}&\oneloop
}$$
which reproduces the previous result up to a constant $C'$. 
The latter may be determined, for example by considering the $s\to 0$ limit, 
and the result reproduces \hciz.
Thus the stationary phase approximation of the original integral 
\hciz\ or \integr\ turns out to give the exact result!
This well known empirical fact
turned out to be a particular case of a general situation analysed
later by Duistermaat and \He\ \DH: if a classical system has only  periodic
trajectories with the same period, the stationary
phase (or saddle point) method is exact. 

In more precise mathematical terms, 
let $\CM$ be a $2n$-dimensional symplectic manifold
with symplectic form $\Go$, and suppose that it is invariant under a
$U(1)$ action. Let  $H$ be the Hamiltonian corresponding
to this action (i.e. $dH= i_V \Go$, $V$ the vector field of
infinitesimal $U(1)$ action). Assume also that the fixed point set 
(the critical points) is discrete. 
 Then the theorem of Duistermaat-\He\ asserts that
the stationary phase method is exact, i.e. that 
\eqn\DHthm{
 \int {\Go^n\over n!} \e{itH}=\({2 \pi \over t}\)^n
\sum_{P_c} 
\e{i {\pi\over 4}{\rm sign}({\rm Hess}(P_c))}\,{\e{it H(P_c)}}
{\sqrt{\det \Go(P_c)}\over
 \sqrt{|\det{\rm Hess}(P_c)| }}
}
where the sum is over (isolated) critical points $P_c$; 
the phase involves 
the signature ${\rm sign}({\rm Hess}(P_c))$, i.e. the number of positive minus 
the number of negative eigenvalues, of the
Hessian matrix ${\rm Hess}_{ij}=\partial^2 H/\partial \xi^i\partial
\xi^j \Big|_{P_c}$. 

The integral \hciz\ satisfies the conditions of the above 
theorem. The integration runs over the orbit $\CO=\{M=UB\Ud\}$ of $B$
under the coadjoint action of $U$: this orbit, 
homeomorphic to  the manifold $U(N)/U(1)^N$, 
has the even dimension $N(N-1)$ and is in fact a symplectic manifold. On two
tangent vectors $V_i=[X_i,M]$, $i=1,2$, ($X_i$ antihermitian),
 tangent to $\CO$ on $M$, the
symplectic form reads $\Go(V_1,V_2)=\Tr M[X_1,X_2]$. 
The Hamiltonian $H=\Tr A M$ 
defines a periodic flow  $M(t)=\e{iAt} M(0) \e{-iAt}$
if all eigenvalues of $A$ are relatively rational.
As the latter configurations form a dense
set among diagonal matrices $A$, this constraint may in fact be 
removed and 
this justifies {\it a posteriori} the stationary phase calculation above. 
For further details on the Duistermaat-\He\ theorem
in the present context, the reader 
may consult also  \refs{\MSt,\Sz}.

%
\newsec{Generalizations and extensions}
\subsec{Other groups}
\nind
As already mentioned, 
the integral \resu\ appeared first in the work of Harish-Chandra \HC,
as a simple application to compact groups of a general discussion
of invariant differential operators on Lie algebras.
Following Harish-Chandra, let $G$ be  a compact Lie group.
Denote by 
${\rm Ad}$ the adjoint action of $G$ on itself and on its Lie algebra $\goth{g}$,
by $\langle .,.\rangle$ the invariant bilinear form on $\goth{g}$,
and by $\goth{h}\subset \goth{g}$ the Cartan subalgebra.
If $h_1$ and $h_2 \in \goth{h}$ 
\eqn\harish{
\Delta(h_1)\Delta(h_2)\int_G \d g \exp \langle{\rm Ad}(g) h_1, h_2\rangle=
{\rm const} \sum_{w\in W} \Ge(w) \exp \langle w(h_1),   h_2\rangle\ ,}
where $w$ is summed over the Weyl group $W$, 
$\Ge(w)=(-1)^{\l(w)}$, $\l(w)$ is 
the number of reflections generating $w$, 
 and $ \Delta(h)=\prod_{\Ga >0} \langle \Ga,h\rangle\ ,$
a product over the positive roots of $\goth{g}$. 
In the case of $U(N)$, if we take $h_1=iA$, $h_2=iB$,  \harish\
 reduces to \resu.

\medskip\nind
This extension to general compact $G$ may also be derived by 
constructing the heat kernel $K(g_1,{\rm Ad}(g)g_2;s)$ on $G$ 
in terms of characters,  by using Weyl's formulae for characters
and by averaging $K$ over $G$~\AI. 
Let's sketch the derivation. One may always assume that
$g_j=e^{h_j}$, $h_j\in \goth{h}$, $j=1,2$ and one writes  
\eqn\ker{
K(g_1,g_2;s)=\sum_\Gl d_\Gl \chi_\Gl(g_1g_2^{-1}) e^{-\oh s C_\Gl}
}
where $C_\Gl$ is the value of quadratic Casimir for the representation of
weight $\Gl$, $d_\Gl$ the dimension of the latter. 
Integration over the adjoint action  gives
\eqn\intK{ \int \d g K(g_1,{\rm Ad}(g)g_2;s)=\sum_\Gl  \chi_\Gl(g_1)
\chi_\Gl(g_2^{-1}) e^{-\oh s C_\Gl}\ .}
Weyl's formula for the characters reads
\eqn\weyl{
\chi_\Gl(e^{ih})={\sum_{w\in W} \Ge_w e^{i\langle\Gl+\rho,
w(h)\rangle}\over 
{\rm same\ for\ }\Gl=0 }
\ ,}
where $\rho$ is the Weyl vector, sum of all fundamental weights.
As $C_{\Gl}=|\Gl+\rho|^2 -|\rho|^2$, the r.h.s of \intK\ is the 
exponential of  a quadratic form in $\Gl$.  
The summation over $\Gl$ may be extended 
from the Weyl chamber to the 
full weight lattice, Poisson summation formula is then used and 
after taking a limit of infinitesimal $s$, $h_1$ and $h_2$, one is  led 
to \harish.

\omit{For example, for the orthogonal group O$(N)$, and for $A$ and
$B$ {\it anti}-symmetric real matrices
$$\int_{O(N)} dO \exp {1\over s} \Tr A O B O^T =
{\cdots\over \cdots }
\ .$$}

For completeness we also mention the generalisation of \hciz\ 
involving Grassmannian coordinates and the integration over
the supergroup $U(N_1|N_2)$, see \refs{\susy,\GW}. 
The case of integration over a 
pseudounitary group $U(N_1,N_2)$ has also been
discussed, see \Fyo\ and further references therein.

\subsec{Rectangular matrices}
\nind Consider the integral
\eqn\rect{I^{(2)}(A,B;s)=\int_{U(N_2)} \D U \int_{U(N_1)} \D V
\exp{N\over s} \, \Tr(AUBV^\dagger +{\rm h.c.})}
where $A$ is a complex $N_1\times N_2$ matrix, $B$ a  complex 
$N_2\times N_1$ matrix, and $N=\min(N_1,N_2)$.
Without loss of generality, one may assume that $N_1\ge N_2$ and 
that the $N_2\times N_2$ matrices $\Ad A$ and $B\Bd$ are diagonal, 
with real non negative eigenvalues $a_i$, $b_i$ respectively. 
We assume again that the $a$'s  on the one hand, the $b$'s on the
other, are all distinct, so that neither of the Vandermonde 
$N_2\times N_2$ determinants $\Delta(a)$, $\Delta(b)$, vanishes.
Then the methods of heat equation or of character expansion 
presented in sec.~2 yield the following expression
\eqn\resrec{
 \Delta(a) \Delta(b) I^{(2)}(A,B;s) = 
{\prod_{p=1}^{N_2 -1} p! \prod_{q=1}^{N_1 -1} q!\over
\prod_{r=1}^{N_1-N_2 -1} r! } (s/N)^{N_2(N_1-1)} {\det I_{N_1-N_2}
(2 N \sqrt{a_i b_j}/s) \over \prod_{i=1}^{N_2} (a_i b_i)^{{1\over2}(N_1-N_2)}}\ , }
with $I_\nu(z)$ the Bessel function 
$ I_\nu(z)=\sum_{n=0}^\infty {1\over n! (n+\nu)!} \(z\over 2\)^{2n+\nu}$.
This expression has been obtained for $N_1=N_2$ 
in \refs{\GW,\Ba} and in the general case in  \JSV.

Integrals \hciz\ and \rect\ are the cases $K=1,2$ of an infinite
set of unitary integrals which are exactly calculable:
\eqnn\genint
$$\eqalignno{
I^{(K)}(A_k,B_k;s)&=\int_{U_k\in U(N_k)} \prod_{k=1}^K \D U_k
\exp{N\over s} \sum_{k=1}^K \Tr A_k U_{k+1} B_k U_k^\dagger\cr
&={\rm const} {\det\phi(a_i b_j(N/s)^K)\over\Delta(a)\Delta(b)}&\genint\cr
}$$
(index $k$ is cyclic modulo $K$)
where $N=\min_k(N_k)$,
$A_k$ and $B_k^\dagger$ are $N_k\times N_{k+1}$ matrices, $1\le k\le K$;
the generalized hypergeometric series $\phi$ is given by
$\phi(x)=\sum_{n=0}^\infty {x^n/\prod_{k=1}^K (n+N_k-N)!}$;
and, assuming $N_1=N$,
the $a_i$ (resp.\ $b_i$) are
the $N$ (distinct) eigenvalues of $A_1 A_2\ldots A_K$
(resp.\ $B_K\ldots B_2 B_1$). 

In what follows, we shall concentrate on integrals \hciz\ and \rect;
however the analysis applies equally well to the more general integral \genint,
see in particular \OS\ for a discussion of hypergeometric tau-functions of
Toda lattice.

\subsec{Correlation functions }
\nind 
Returning to the unitary group and the integral \hciz, 
it is also natural to consider the correlation functions associated 
with it, i.e. to compute the integrals
\eqn\correl{
\int_{U(N)} \D U\, U_{i_1j_1}\cdots U_{i_mj_m}\Ud_{k_1\l_1}\cdots
\Ud_{k_m\l_m}\, \exp{N\over s}\Tr AUB\Ud \ .
}
Partial results have been obtained in \Shat\ 
and in  \Mor. 
We still lack explicit and general expressions for these
correlation functions.

\def\L{\ell}
\def\Lb{\bar{\ell}}
\def\M{m}
\def\Mb{\bar{m}}
\def\X{g}
\def\Y{h}
\def\A{a}
\def\B{b}
\def\P{\varphi}
\def\Pb{\bar{\varphi}}

%
\newsec{Connection with integrable hierarchies}
\nind The  integral \hciz\ turns out to provide a non-trivial solution
of the two-dimensional Toda lattice hierarchy \PZJ. This stems
from the following observation: define
\eqn\deftau{
\tau_N=
{\det\big(\E{{1\over\hbar} a_i b_j}\big)_{1\le i,j\le N} 
\over \Delta(a)\,\Delta(b)\ }
}
where $\hbar=s/N$. Comparing with Eq.~\resu,
we see that $\tau_N=\hbar^{-N(N-1)/2} \prod_{p=0}^{N-1} (p!)^{-1}
I$. Then the following formula holds: (see also \refs{\PZJ,\KM} for
a two-matrix-model formulation)
\eqn\detform{
\tau_N=
\det\left(\oint\!\!\oint {\d u\over 2\pi i u}{\d v\over 2\pi i v}
u^j v^i \E{{1\over\hbar}\left(\sum_{q\ge 1} t_q u^q+\sum_{q\ge 1} \bar{t}_q v^q+u^{-1}v^{-1}\right)}
\right)_{0\le i,j\le N-1}
}
where the integration contours are small enough circles around
the origin, and with the traditional notations:
\eqn\deftq{
t_q= \hbar {1\over q} \sum_{i=1}^N a_i^q={s\over q} \s_q
\qquad
\bar{t}_q = \hbar {1\over q} \sum_{i=1}^N b_i^q={s\over q}\t_q
\qquad q\ge 1
}
Formula \detform\ can be easily proved by noting that
$\e{{1\over\hbar}\sum_q t_q u^q}=\prod_{i=1}^N (1-ua_i)^{-1}$,
$\e{{1\over\hbar}\sum_q \bar{t}_q v^q}=\prod_{i=1}^N (1-vb_i)^{-1}$,
and expanding the contours to catch the poles at $u=a_i^{-1}$,
$v=b_i^{-1}$. This makes $\tau_N$ a tau function of the 2D Toda
lattice hierarchy, as we shall discuss now.

\subsec{Biorthogonal polynomials and 2D Toda lattice hierarchy}
\noindent Noting that the parameter
$\hbar=s/N$ can always be scaled away we set $\hbar=1$
throughout this section. We take Eq.~\detform\ 
to be the
definition of $\tau_N$ as a function of the two infinite sets
of times $( t_q ,  \bar{t}_q )$, $q\ge 1$. 
We also set $\tau_0=1$.

Formula \detform\ suggests to introduce a
non-degenerate bilinear form on the space of polynomials by
\eqn\bil{
\braket{q}{p}=
\oint\!\!\oint {\d u\over 2 \pi i u}{\d v\over 2 \pi i v}
p(u) q(v)
\e{\sum_{q\ge 1} t_q u^q+\sum_{q\ge 1} \bar{t}_q v^q+u^{-1}v^{-1}}
}
and normalized biorthogonal polynomials $q_n(v)$ and $p_n(u)$ 
with respect to the bilinear form above,
that is polynomials of the form
$p_n(u)=h_n^{-1} u^n+\cdots$ and $q_n(v)=h_n^{-1} v^n+\cdots$, such that
$\braket{q_m}{p_n}=\delta_{mn}$ for all $m,n\ge 1$. 
One can now replace monomials
in Eq.~\detform\ with biorthogonal polynomials and obtain
immediately
\eqn\tauh{
\tau_N=\prod_{i=0}^{N-1} h_i^2\ .
}

Next we introduce the matrices of multiplication by $u$ and $v$
in the basis of biorthogonal polynomials:
\eqn\LL{
L_{mn}=\bra{q_m} u \ket{p_n}\qquad \bar{L}_{mn}=\bra{q_m} v \ket{p_n}
\qquad m,n\ge 1\ .
}
Note that
\eqn\LLb{
L_{mn}=0\quad m>n+1\ , \qquad \bar{L}_{mn}=0\quad n>m+1\ .
}

A standard calculation \PZJ\ leads to the following evolution equations
for $L$ and $\bar{L}$ with respect to variations of the $t_q$, $\bar{t}_q$:
\eqna\lax
$$\eqalignno{
{\der L\over\der t_q}=-[(L^q)_+,L]\qquad&\qquad
{\der L\over\der \bar{t}_q}=[(\bar{L}^q)_-,L]
&\lax{a}\cr
{\der \bar{L}\over\der t_q}=-[(L^q)_+,\bar{L}]
\qquad&\qquad
{\der \bar{L}\over\der \bar{t}_q}=[(\bar{L}^q)_-,\bar{L}]\ .&\lax{b}\cr
}$$
Here $(\cdot)_\pm$ denotes the lower / upper triangular part plus
one half of the diagonal part. Eqs.~\lax{} are the standard form of
the two-dimensional Toda lattice hierarchy \UT\ (up to a choice of
sign of the $t_q$).

It is often more convenient to write these equations as
quadratic equations in the set of $(\tau_N)$. These are
the ``bilinear'' Hirota equations. They are obtained by
picking two sets of times $( x_q, \bar{x}_q)$ and
$(y_q, \bar{y}_q)$ and
writing in two different ways
$$
\oint\!\!\oint {\d u\over 2 \pi i u}{\d v\over 2 \pi i v}
p_{n,x,\bar{x}}(u) q_{m,y,\bar{y}}(v)
\e{\sum_{q\ge 1} x_q u^q+\sum_{q\ge 1} \bar{y}_q v^q+u^{-1}v^{-1}}
$$
where $p_{n,x,\bar{x}}$ is the right biorthogonal polynomial associated to times
$(x_q,\bar{x}_q)$, whereas 
$q_{m,y,\bar{y}}$ is the left biorthogonal polynomial associated
to times $(y_q,\bar{y}_q)$.
One finds:
\eqnn\hirota
$$
\eqalignno{
&\oint {\d u\over 2\pi i u} u^{n-m}
\tau_n(x_q-{1\over q} u^{-q},\bar{x}_q)
\tau_{m+1}(y_q+{1\over q} u^{-q},\bar{y}_q)
\e{\sum_{q\ge 1}(x_q-y_q) u^q}\cr
=&\oint {\d v\over 2\pi i v} v^{m-n}
\tau_m(y_q,\bar{y}_q-{1\over q}v^{-q})
\tau_{n+1}(x_q,\bar{x}_q+{1\over q}v^{-q})
\e{\sum_{q\ge 1}(\bar{y}_q-\bar{x}_q) v^q}\ .&\hirota\cr
}
$$
Equivalently, equation \hirota\ can be derived directly from 
the original expression \deftau\ using the set of determinant identities
\eqn\detid{
\sum_{i=1}^{n+1} \det(A-a_i,B) \det(A'+a_i,B')=\sum_{j=1}^{m+1} \det(A',B'-b'_j)\det(A,B+b'_j)
}
where $A=\{a_1,\ldots,a_{n+1}\}$, $B=\{b_1,\ldots,b_n\}$,
$A'=\{a'_1,\ldots,a'_m\}$, $B'=\{b'_1,\ldots,b'_{m+1}\}$, and a symbolic notation $\det(\cdot,\cdot)$ is used
for determinants of the form $\det(\exp({1\over\hbar} a_i b_j))$.

Expanding Eq.~\hirota\ in powers of $x_q-y_q$ and $\bar{x}_q-\bar{y}_q$, expanding in power series
in $u$, $v$ and performing the integration results
in an infinite set of partial differential equations satisfied
by the $(\tau_N)$.

{\it Example:} Expand to first order in $x_1-y_1$, and set $m=n+1$.
The result is:
\eqn\jac{
\tau_{n+1}\tau_{n-1}=\tau_n \der\bar{\der} \tau_n-\der\tau_n
\bar{\der}\tau_n\qquad \forall n\ge 1
}
(with $\der=\der/\der t_1$, $\bar{\der}=\der/\der\bar{t}_1$)
which is a form of the Toda lattice equation.
It is of course also the Desnanot--Jacobi determinant identity
applied to Eq.~\detform.

Finally, the matrices $L$ and $\bar{L}$ satisfy an additional relation,
the so-called {\it string equation}, which takes the form \PZJ
\eqn\se{
[L^{-1},\bar{L}^{-1}]=1\ .
}

\subsec{Large $N$ limit as dispersionless limit}
\nind We now restore the parameter $\hbar$, which is required for the
large $N$ limit. Indeed, as $N\to\infty$, $\hbar$ must be sent to zero,
keeping $s=\hbar N$ fixed. We define 
\eqn\defdisp{
f(t_q,\bar{t}_q;s)=\lim_{N\to\infty} \hbar^2 \log \tau_N(t_q,\bar{t}_q)
}
where $\tau_N(t_q,\bar{t}_q)$ is defined by Eq.~\detform.
In a region of the space of parameters
$(t_q,\bar{t}_q)$
which includes a neighborhood of the origin $t_q=\bar{t}_q=0$,
$f(t_q,\bar{t}_q;s)$ is governed by ``dispersionless'' equations we
shall describe now. $f(t_q,\bar{t}_q;s)$ is called in this context
the dispersionless tau-function.

First, comparing with the expression \freeen\ of the free energy 
$F(\s_q s^{-q/2},\t_q s^{-q/2})$
(where we have explicitly stated the dependence of $F$
on the $\s_q$ and the $\t_q$ and scaled away the parameter $s$),
using Eq.~\resu\ and definitions \deftq, we see that
\eqn\dispfree{
f(t_q,\bar{t}_q;s)=-{1\over 2} s^2 \log s +{3\over 4}s^2
+s^2 F(q\,t_q s^{-q/2-1},q\,\bar{t}_q s^{-q/2-1})\ .
}
This is the
scaling form of the dispersionless tau function.\foot{
In fact, one can 
get rid of one more parameter since $A$ and $B$ can be
scaled independently, but we choose not to do so
for symmetry reasons.}

The dispersionless Toda hierarchy is a classical limit of
the Toda hierarchy: it is obtained by replacing commutators
with Poisson brackets defined by
\eqn\bracket{\{ \X(z,s), \Y(z,s) \}
= z{\der \X\over\der z}  \, {\der\Y\over\der s}
-z{\der\Y\over\der z} \, {\der\X\over\der s}
}
Here $z$ is the classical analogue of the shift operator 
$Z_{ij}=\delta_{i\,j+1}$, which justifies that $\{ \log z,s\}=1$.

The Lax Eqs.~\lax{} become
\eqna\displax
$$\eqalignno{
{\der \L\over\der t_q}=-\{(\L^q)_+,\L\}\qquad&\qquad
{\der \L\over\der \bar{t}_q}=\{(\Lb^q)_-,\L\}
&\displax{a}\cr
{\der \Lb\over\der t_q}=-\{(\L^q)_+,\Lb\}
\qquad&\qquad
{\der \Lb\over\der \bar{t}_q}=
\{(\Lb^q)_-,\Lb\}&\displax{b}\cr
}$$
where $(\cdot)_\pm$ now refers to positive and negative parts of
Laurent expansion in $z$, and
$\L(z,s,t)$ and $\Lb(z,s,t)$ have the following $z$ dependence:
\eqna\Ldisp
$$\eqalignno{
\L&=r\, z+\sum_{k=0}^\infty \lambda_k\, z^{-k}&\Ldisp{a}\cr
\Lb&=r\, z^{-1}+\sum_{k=0}^\infty \bar{\lambda}_k\, z^k&\Ldisp{b}\cr
}$$
related to the structure \LLb\ of $L$ and $\bar{L}$.

These equations imply that the differential forms $\d\P(\L,s,t)$,
$\d\Pb(\Lb,s,t)$ are closed:
\eqna\closed
$$\eqalignno{
\d\P&=\M\, \d \L/\L
+\log z\, \d s+\sum_{q\ge 1} (\L^q)_+ \d t_q 
+\sum_{q\ge 1} (\Lb^q)_- \d \bar{t}_q &\closed{a}\cr
\d\Pb&=\Mb\, \d \Lb/\Lb
+\log z\, \d s+\sum_{q\ge 1} (\L^q)_+ \d t_q 
+\sum_{q\ge 1} (\Lb^q)_- \d \bar{t}_q &\closed{b}\cr
}$$
where $\M$ and $\Mb$ are Orlov--Shulman functions,
which can be characterized by
\eqna\defM
$$\eqalignno{
\M&=\sum_{q\ge 1} q t_q\, \L^q + s + \sum_{q\ge 1}
{\der f\over\der t_q} \, \L^{-q}&\defM{a}\cr
\Mb&=\sum_{q\ge 1} q \bar{t}_q\, \Lb^q + s + \sum_{q\ge 1}
{\der f\over\der \bar{t}_q} \, \Lb^{-q}&\defM{b}\cr
}$$
and satisfy the ``dressed'' Poisson bracket relations
\eqn\ML{
\{ \log \L, \M \} = 1 \qquad \{ \log \Lb, \Mb \} = 1\ .
}

In the present case, we have the following
constraints, which determine uniquely the solution of 
the dispersionless Toda hierarchy:
\eqn\const{
\M=\Mb= (\L\Lb\,)^{-1}\ .}
Eqs.~\const\ are directly 
related (via Eqs.~\ML) to the classical limit of the string equation~\se,
i.e. $\{ \L^{-1},\Lb^{-1} \}=1$. 

Let us call $\A=\L^{-1}$ and $\B=\Lb^{-1}$.
The fact that $\M=\Mb$
implies\foot{%
The constraint $\M=\Mb$ is true for a more general class of solutions of Toda,
see for example sec.~4.4 and \Ivan.}
that $\P(\A)$ and $\Pb(\B)$
are related by Legendre transform, or that
their derivatives $\B(\A)={\d\over\d \A} \P(\A)$
and $\A(\B)={\d\over\d \B} \Pb(\B)$
are functional inverses of each other
(the latter fact was derived by {\it ad hoc}\/ methods
in \Maty\ and \PZJb).

\subsec{Application of the dispersionless formalism}
\nind In what follows we set $s=1$, so that $\theta_q=q t_q$. 
In order to explore the structure of the function $f(t_q,\bar t_q;s)$, we
 now assume that only a finite number of $t_q$ and $\bar{t}_q$ is
non-zero. Note that this cannot happen if the eigenvalues are real;
however, here we are interested in properties of the  integral \hciz\ 
as a formal power series in the $t_q$ and therefore we do not worry about the
actual support of the eigenvalues.
Then, according to Eqs.~\Ldisp{} and \defM{},
the Laurent expansion of $\A(z)$ and $\B(z)$ is
finite, of the form:
\eqn\finwid{
\A=\sum_{q=1}^{\bar{n}+1} \alpha_q z^{-q}
\qquad \B=\sum_{q=1}^{n+1} \beta_q z^q
}
where $n=\max\{q|t_q\ne 0\}$,
$\bar{n}=\max\{q|\bar{t}_q\ne 0\}$, $\alpha_1=\beta_1=1/r$,
$\alpha_{\bar{n}+1}=\bar{n} \bar{t}_{\bar{n}} r^{\bar{n}+1}$,
$\beta_{n+1}=n t_n r^{n+1}$ ($r$ is defined by Eqs.~\Ldisp{}).
Then it is easy to show that $\A$ and $\B$ satisfy an algebraic equation,
some coefficients of which can be worked out explicitly:
\eqn\algeq{
b^{\bar{n}+1} a^{n+1}
-b^{\bar{n}} a^{n}
-b^{\bar{n}}(a^{n-1} \s_1+\cdots+\s_{n})
-a^n(b^{\bar{n}-1} \t_1+\cdots+\t_{\bar{n}})
+{\scriptstyle\rm lower\ order\ terms}=0\ .
}
Plugging Eq.~\finwid\ into Eq.~\algeq\ leads to algebraic equations
for the coefficients $\alpha_q$, $\beta_q$ as a function of
the $\s_q$, $\t_q$.

{\it Examples:}

$\bullet$ If $\t_q=\delta_{q1}\t_1$, one can set $\t_1=1$ by homogeneity.
In this case Eq.~\algeq\ is quadratic in $b$.
In terms of $b$ and $\ell=1/a$ it reads 
\eqn\ex{
b^2-b(\ell+\theta_1 \ell^2+\cdots
+\theta_n \ell^{n+1})-\ell P(\ell)=0
}
where $P$ is a polynomial with $P(0)=1$.
One way to determine $P$ is to note that Eq.~\finwid\ 
provides a rational parameterization of $a$ and $b$, so that
the resulting curve $(a,b)$ has genus zero, which forces the discriminant
of Eq.~\ex\ to be 
of the form $\ell(1+\ell/ (4\psi^2))Q(\ell)^2$ 
where $Q$ is a polynomial of degree $n$ with $Q(0)=2$, and
$\psi$ is a constant.
This yields the expression
\eqn\exa{
b(\ell)={1\over 2}\left(\ell+\theta_1 \ell^2+\cdots+\theta_n \ell^{n+1}
+\sqrt{\ell\Big(1+{\ell\over 4\psi^2}\Big)} Q(\ell)\right)\ .
}
Considering Eq.~\defM{a} as an asymptotic expansion of $\M=b/\ell$ as
$\ell\to\infty$ fixes $Q$: 
$Q$ is the polynomial part of $2\psi(1+\theta_1 \ell+\cdots+
\theta_n\ell^n)/\sqrt{1+4\psi^2/\ell}$. Imposing $Q(0)=2$ leads to
\eqn\exb{
\psi=1+\sum_{q=1}^n (-1)^{q+1}  {(2q)!\over (q!)^2} \theta_q \psi^{2q+1}
}
Note that $\psi=r^2={\der^2 F/\der\theta_1\der\bar{\theta}_1}
=\exp(\der^2 F/\der s^2)$ (the latter equality being the
dispersionless limit of Toda Eq.~\jac).

By Lagrange inversion, Eq.~\exb\ allows to get the exact expansion of 
$\psi$ as well as of
$F$ in powers of the $\theta_q$. One finds, restoring the $\t_1$ dependence:
\eqnn\exc
$$\eqalignno{
\psi&=\sum_{\alpha_1,\ldots,\alpha_n=0}^\infty 
{(\sum_{q\ge 1} (2q+1)\alpha_q)!\over (\sum_{q\ge1}2 q\alpha_q+1)!}
\prod_{q\ge 1}{1\over\alpha_q!} 
\left((-1)^{q+1} {(2q)!\over(q!)^2} \theta_q\right)^{\alpha_q}
\t_1^{\sum_q q \Ga_q}
 \cr
F&=\sum_{\alpha_1,\ldots,\alpha_n=0}^\infty 
{(\sum_{q\ge 1} (2q+1)\alpha_q-3)!\over (\sum_{q\ge1}2 q\alpha_q)!}
\prod_{q\ge 1}{1\over\alpha_q!} 
\left((-1)^{q+1} {(2q)!\over(q!)^2} \theta_q\right)^{\alpha_q}
\t_1^{\sum_q q \Ga_q} 
\ .
&\exc\cr
}
$$

$\bullet$ If $\theta_q=\delta_{qn} \theta_n$, $\bar{\theta}_q=\delta_{qn}
\bar{\theta}_n$ (with the same $n$),
one can show using the formalism above that
$\psi={\der^2 F/\der\theta_1\der\bar{\theta}_1}$ satisfies the equation:
\eqn\exd{
\psi=1+(n+1) \theta_n \bar{\theta}_n \psi^{n+2}
}
so that
\eqn\exe{
\psi=\sum_{k\ge 1} (n+1)^k {((n+2)k)!\over((n+1)k+1)!k!}
(\theta_n \bar{\theta}_n)^k
\qquad
F=\sum_{k\ge 1} (n+1)^k {((n+2)k-3)!\over((n+1)k)!k!}
(\theta_n \bar{\theta}_n)^k\ .
}
For example, 
for $n=2$, 
\def\thd{(\s_2\t_2)}
$F=\oh \thd +{3\over 4} \thd^2 +{9\over 2} \thd^3+\cdots$

\subsec{Case of rectangular matrices}
\noindent
All of the formalism above applies equally well to the integral \rect\ over
rectangular matrices. The latter has a ``diagonal'' character expansion i.e.
\eqn\diagchar{
\tau_N^{(2)}={\rm const}\,I^{(2)}(A,B;s)=
\sum_{\Gl} \hbar^{-|\Gl|} {1\over \prod_p (\lambda_p+p-1)!
(\lambda_p+p-1+\nu)!} \,{\Delta_\lambda(a)\over \Delta(a)}
{\Delta_\lambda(b)\over \Delta(b)}
}
where $\hbar=s/N$, $N=\min(N_1,N_2)$, 
$\nu=|N_1-N_2|$.

The form of Eq.~\diagchar\ 
alone proves that the $(\tau_N^{(2)})$ form a tau function of 2D Toda lattice
(for fixed $\nu$).
There is also a determinant formula similar to Eq.~\detform\ 
which we shall not write.

The large $N$ limit is taken keeping 
$\hbar N=s$ and $\hbar \nu\equiv \xi$ fixed.
The times are $t_q=\hbar{1\over q} \Tr (A A^\dagger)^q$, $\bar{t}_q=
\hbar{1\over q} \Tr (B B^\dagger)^q$.
Biorthogonal polynomials and the dispersionless equations can
be obtained similarly as above.

We have thus constructed a whole family 
$g_\xi(t_q,\bar{t}_q;s)=\lim_{N\to\infty} \hbar^2 \log \tau_N^{(2)}$ of
dispersionless tau functions.
The diagonal character expansion \diagchar\ implies
that they belong to the class
of solutions that satisfy $\M=\Mb$. After some calculations,
one finds that the constraint analogous to Eq.~\const\ is
\eqn\rectconst{
\M(\M+\xi)=\Mb(\Mb+\xi)=(\L\Lb\,)^{-1}
}

Note that if $\xi=0$ i.e.\ $N_1=N_2$, the constraint \rectconst\ becomes
$\M=\Mb=(\L\Lb\,)^{-1/2}$,
which suggests to redefine $\L=\L_2^{\,2}$,
$\Lb=\Lb_2^{\,2}$. The new functions $\L_2$, $\Lb_2$ satisfy the
string equation $\{ \L_2^{-1},
\Lb_2^{-1} \}=1$ of the integral \hciz, and of course the same
Toda equations \displax{} for even times, with the replacement
$t_{2q}\to t_q$, $\bar{t}_{2q}\to \bar{t}_q$. In fact, one can check that
we recover in this case the dispersionless
tau function of the usual integral \hciz\ with odd moments equal to zero:
$g_0(u_q,\bar{u}_q;s)
=2 f(t_{2q}=u_q,t_{2q+1}=0,\bar{t}_{2q}=\bar{u}_q,
\bar{t}_{2q+1}=0;s)$.

%
\newsec{Diagrammatic expansion and combinatorics} 
\subsec{A Feynman diagram expansion of $F$}
\noindent
To develop a Feynman diagram expansion of $\hciz$, we first trade the 
integration over the unitary group for an integration over $N\times N$
complex matrices  $X$ by writing
\eqn\Xint{
\e{{N\over s} \Tr AUB\Ud}=\big({\pi\over N}\big)^{-N^2}
\int \D X \D\Xd \e{-N \Tr X\Xd}
\exp{{N\over \sqrt{s}} \Tr(AU\Xd+XB\Ud)}} 
Thus,
\eqn\newI{I(A,B;s)={\rm const}\int \D X \D \Xd \e{-N\Tr (X\Xd)
+N^2 W(AXB\Xd;s)}
}
where 
\eqn\defW{\e{ N^2 W(X_1X_2;s)}=
\int \D U \exp {N\over\sqrt{s}}\, 
\Tr (UX_1+\Ud X_2)\ .}
This integral is known exactly in this large $N$ limit \OBZ:
with the same abuse of notations as mentionned in sec. 1,
\eqna\Wresu
$$\eqalignno{W(Y;s)&= \sum_{n=1}^\infty s^{-n} \sum_{\Ga\,\vdash n}
W_\Ga\, {\tr_\Ga Y \over  \prod_p\(\Ga_p! \, p^{\Ga_p}\)} &\Wresu{a}\cr
W_\Ga&= (-1)^n {(2n+\sum \Ga_p -3)!\over (2n)!}
\prod_{p=1}^n \({-(2p)!\over p! (p-1)!}\)^{\Ga_p}\ ,&\Wresu{b}\cr 
}$$
and $W_\Ga$ is an integer, as follows from the recursion formulae
discussed in \OBZ.

The form \newI\ is adequate to develop a diagrammatic expansion \`a la
Feynman of $F(A,B;s)$ (see \refs{\BIZ,\Zv} for reviews).
Recall that double lines are conveniently introduced to encode the 
conservation of indices \tH.
The inverse of the first term in the exponential of \newI\ yields the
propagator,  
$\langle X_{ij} \Xd_{kl}\rangle= \propagarr= 
{1\over N} \delta_{i\ell}\delta_{jk}$,
while each term in $W$, i.e. each monomial 
$\tr_\Ga(AXB\Xd)/ \prod_p\(\Ga_p! \, p^{\Ga_p}\)$ 
gives rise to a  multi-vertex of 
type $\Ga$, (see Fig.~1),  which comes with a weight 
${N^{2-\sum \Ga_p}} W_\Ga $ times products of $a$'s and 
$b$'s and Kronecker symbols expressing the conservation of indices.
When these Kronecker symbols are ``contracted'', they leave 
sums over closed circuits (or faces) of powers of $a$'s or $b$'s. 
A face of side $p$ thus contributes a factor $N \s_p$, resp.\ $N \t_q$
with the notation of \symmfn. To put it another way, 
the Feynman diagrams may be regarded as bicolourable, 
with faces carrying alternatingly $a$ or $b$ ``colour''.

As is well known, only connected diagrams contribute to the free 
energy $\log I$. In the discussion of this connectivity, each 
multi-vertex just introduced must be regarded as a connected object, 
and it is useful to keep track of this fact by drawing a tree
of dotted lines which connect the various traces which compose it, 
see Fig.~1.b.

\hfig{(a) a vertex of type $[2]$; (b) a multi-vertex of type $[1^2\,2^1\,3^1]$.
}{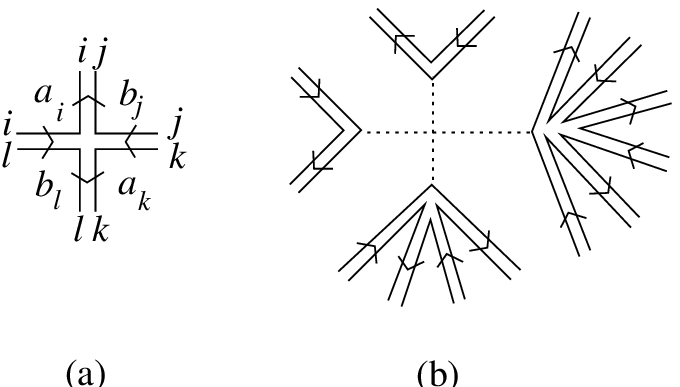}{6cm}

If a diagram contributing to $\log I$ has 
$P$ propagators, 
$V_\Ga$ vertices of type $\Ga$ and $L$ 
loops of indices, and  builds a surface of total genus $g$ with 
 $c$  connected components, 
 it carries a power of $N$ equal to
$$\eqalign{\#(N) &=-P+\sum_\Ga V_\Ga (2-\sum_p \Ga_p) +L\cr &=
(\sum_\Ga V_\Ga\sum_p\Ga_p -P+L) +2 \sum_\Ga V_\Ga (1-\sum_p \Ga_p)\cr &=
2-2g +2\Big(c-1- \sum_\Ga V_\Ga (\sum_p \Ga_p -1)\Big)\le 2-2g \ ,}$$
where the last inequality expresses that to separate the diagram into  
$c$ connected components, one must cut at most its 
$\sum_\Ga V_\Ga (\sum_p \Ga_p -1)$ dotted lines.

This simple counting (see the second ref \OBZ)
shows that the leading $O(N^2)$ terms contributing to $F$
are obtained as the sum of planar (i.e. genus $g=0$) and minimally
connected Feynman diagrams, such that cutting any dotted line
makes them disconnected. They are thus trees in these dotted lines.

\omit{The sign attached to each diagram is 
$(-1)^L=(-1)^{{\rm power\ of\ }\s +\ {\rm  power\ of\ }\t} $.}

These Feynman rules, supplemented by the usual prescriptions
for the symmetry factor of each diagram (equal to the inverse 
of the order of its automorphism group), are what is needed to
get a systematic expansion of $F$ in powers of the $\s$ and $\t$. 

\eqn\Feynm
{F=\sum_{{\rm minimally\ connected}\atop{\rm diagrams}\ \CG}
{s^{-P}  
\over |{\rm Aut} \, \CG|}
\prod_{{\rm vertices}} 
W_\Ga  \prod_{a-{\rm faces}} \s_p \prod_{b-{\rm faces}} \t_q \ .}

Note that these rules imply that the sign attached to each diagram
(coming from the signs of the $W_\Ga$)
is also $(-1)^L=(-1)^{{\rm power\ of\ }\s +\ {\rm  power\ of\ }\t} $.

\bigskip

{\it Examples:}\par\nobreak
$W(X\Xd;s)$ contains a term linear in $X\Xd$ (with coefficient
$1/s$), which may be inserted in arbitrary number on any 
propagator without changing the topology of the diagram: 
each individual insertion (which raises the 
corresponding power of $a$ and $b$ by one unit) is depicted 
by a cross on the propagator.

Terms in $F$ linear in $\s$ and $\t$
come solely from these insertions, which yields 
$\sum_{p=1}^\infty {1\over p} {\s_p \t_p\over s^p}$:
{\epsfxsize=10mm\hbox{\raise -4mm\hbox{\epsfbox{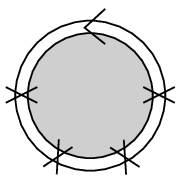}}}}
(a single loop with $p$ insertions has a symmetry factor equal to $p$).

Terms in $F$ of the form $(\s_1\t_1/s)^p$ come entirely from 
a multivertex of type $\tr_{[1^p]} X\Xd$, whence 
a contribution $ 2^p {(3p-3)!\over p!(2p)!}$
{\qquad \epsfxsize=20mm\hbox{\raise -10mm\hbox{\epsfbox{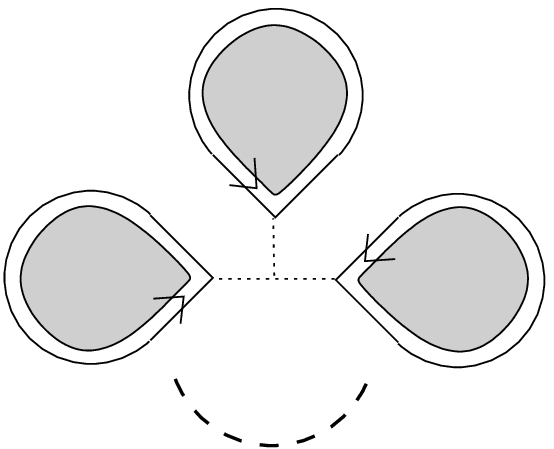}}}}.
More generally, the term in $F$ of the form $\prod_p(\s_p)^{\Ga_p}\t_1^n$ 
comes solely from the multivertex of type $\Ga$, $\Ga\vdash n$, 
in which the contraction of $b$ lines into ``petals'' is unique and 
determines that of $a$ lines.
The weight $W_\Ga$ together 
with the symmetry factor $\prod_p p^{\Ga_p} \Ga_p! $ reproduces the
result of \exc. 

Likewise, it is easy to find the diagrams contributing to 
the $(\s_2\t_2/s^2)^2$ and $(\s_2\t_2/s^2)^3$ terms in $F$
{\qquad \epsfxsize=30mm\hbox{\raise -10mm\hbox{\epsfbox{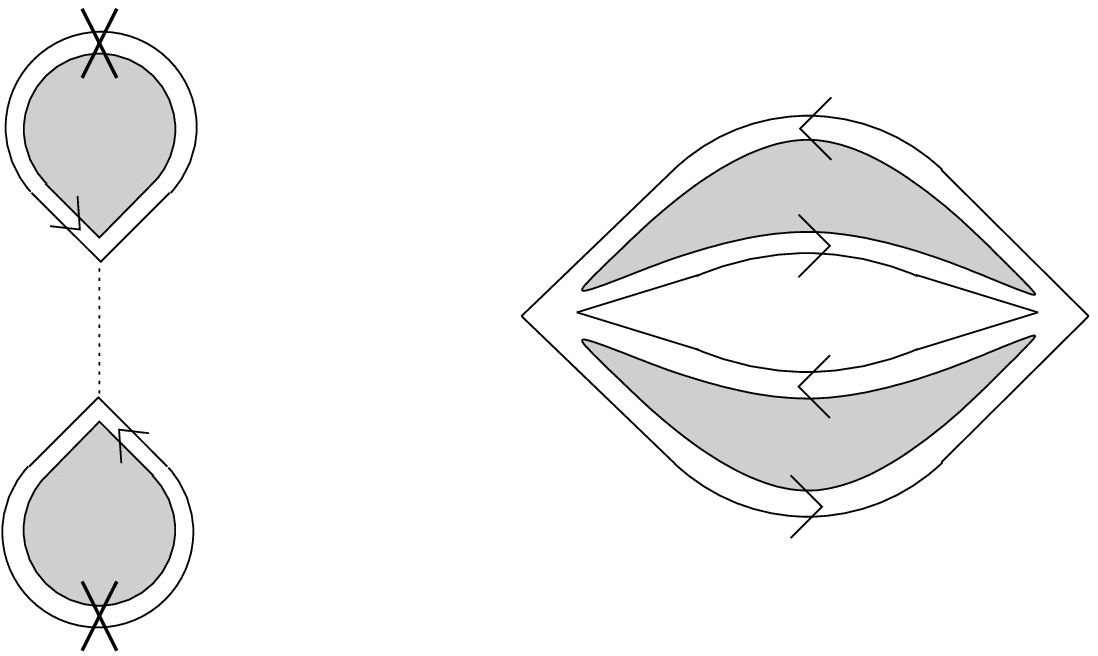}}}}
which give $(\oh +(\oh)^2 )(\s_2\t_2/s^2)^2 
={3\over 4}(\s_2\t_2/s^2)^2$ and 

\noindent 
{\quad \epsfxsize=50mm\hbox{\raise -10mm\hbox{\epsfbox{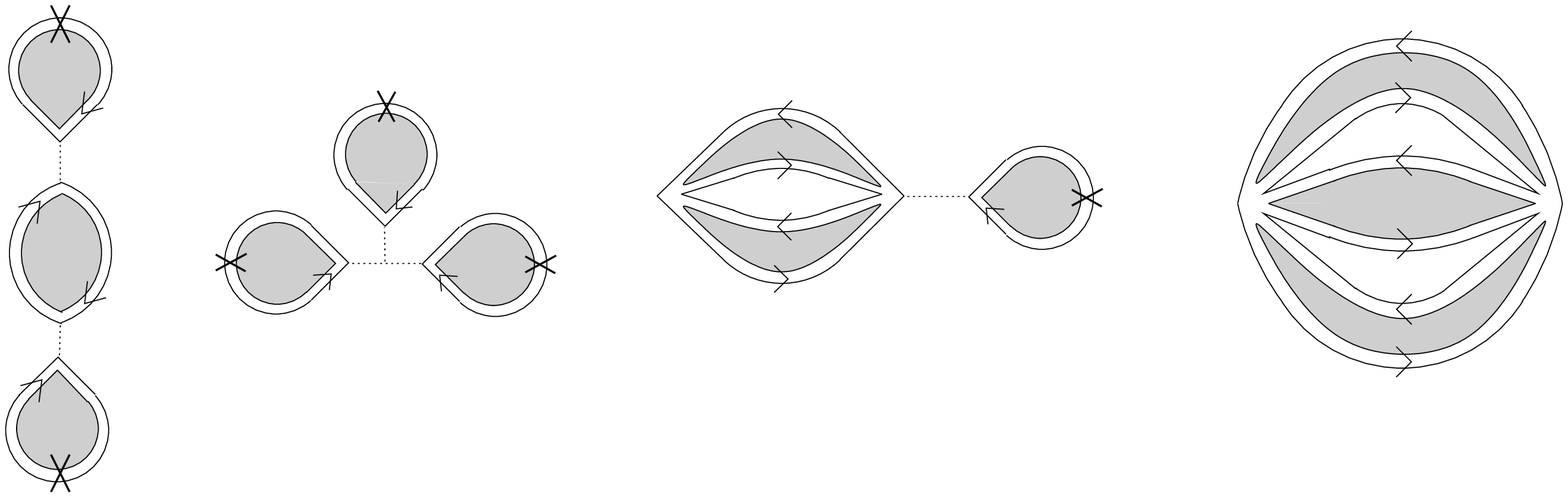}}}}
: \ $(\oh+ {1\over 3!}\times 8+ \oh\times 4 +{1\over 3!}\times 2^2)
\({\s_2\t_2\over s^2}\)^3={9\over 2}\({\s_2\t_2\over s^2}\)^3
$, in agreement with the expansion of the end of sec.~4.
Conversely, one sees how effective the methods of sec.~4 are to
resum classes of Feynman diagrams. Indeed it is not obvious how
to derive directly from the diagrammatic expansion
that $F(\s_2\t_2)$, or more generally
$F(\s_n\t_n)$ and $\psi(\s_n\t_n)$, have the simple form given by Eq.~\exe.
In particular it would be interesting to find a direct combinatorial proof
of the equation \exd\ satisfied by $\psi$. Its form suggests a possible
connection with decorated rooted trees, perhaps in the spirit of \BMS.

{\it Remark}: Actions of the form of Eq.~\newI\ and their diagrammatic 
expansion are a generalization of the ``dually weighted graphs'' of \KSW:
the extra ingredient is that graphs are 
required to be bicolored and the black/white faces are weighted
separately. They are also a generalization
of the bicolored diagrams of \PDF\ -- the faces of the 
latter being weighted only
according to their color and not
to their number of edges (this corresponds to the case where
$A$ and $B$ are projectors).

\subsec{Combinatorics}
\nind We now show that this expansion is in fact equivalent to the one
recently obtained in \BC. What is remarkable is that the methods are
orthogonal: \BC\ is based on manipulations in the symmetric group,
and in particular uses some results on the number of solutions of equations
for permutations (which are however
themselves related to planar constellations \BMS).
We now explain the results of \BC\ that are relevant here, and show their
equivalence to our Feynman diagram expansion.

First we explain how to associate to a (not necessarily connected)
bicolored map $g$, a pair of permutations $\sigma(g)$
and $\tau(g)$ of the set of edges. 
$\sigma(g)$ (resp.\ $\tau(g)$) is the permutation
which to an edge associates the next edge obtained by clockwise rotation
around the white (resp.\ black) vertex to which it is connected, see the
example of Fig.~2a). This is a one-to-one correspondence.

An important remark is that the permutation $\sigma(g)\tau(g)$ encodes
the {\it faces}\/ of the map. Indeed, 
cycles of $\sigma(g)\tau(g)$ are obtained by turning clockwise around each face
and keeping only the edges going from a white to a black vertex.
It is implied that each connected component has its own face at infinity
(for which rotation must be made counterclockwise).
Consequently, the lengths of the cycles of $\sigma(g)\tau(g)$
are precisely one half of the sizes of the faces.
\fig{a) A bicolored planar map and b) its dual.
$\sigma=\hbox{(1 2 11) (3 4 5) (6 7 10) (8 9)}$,
$\tau=\hbox{(1) (2 3 6) (4 5) (7 8) (9 11 10)}$,
$\sigma\tau=\hbox{(1 3 5 6 8 11) (2 10) (4) (7 9)}$.
}{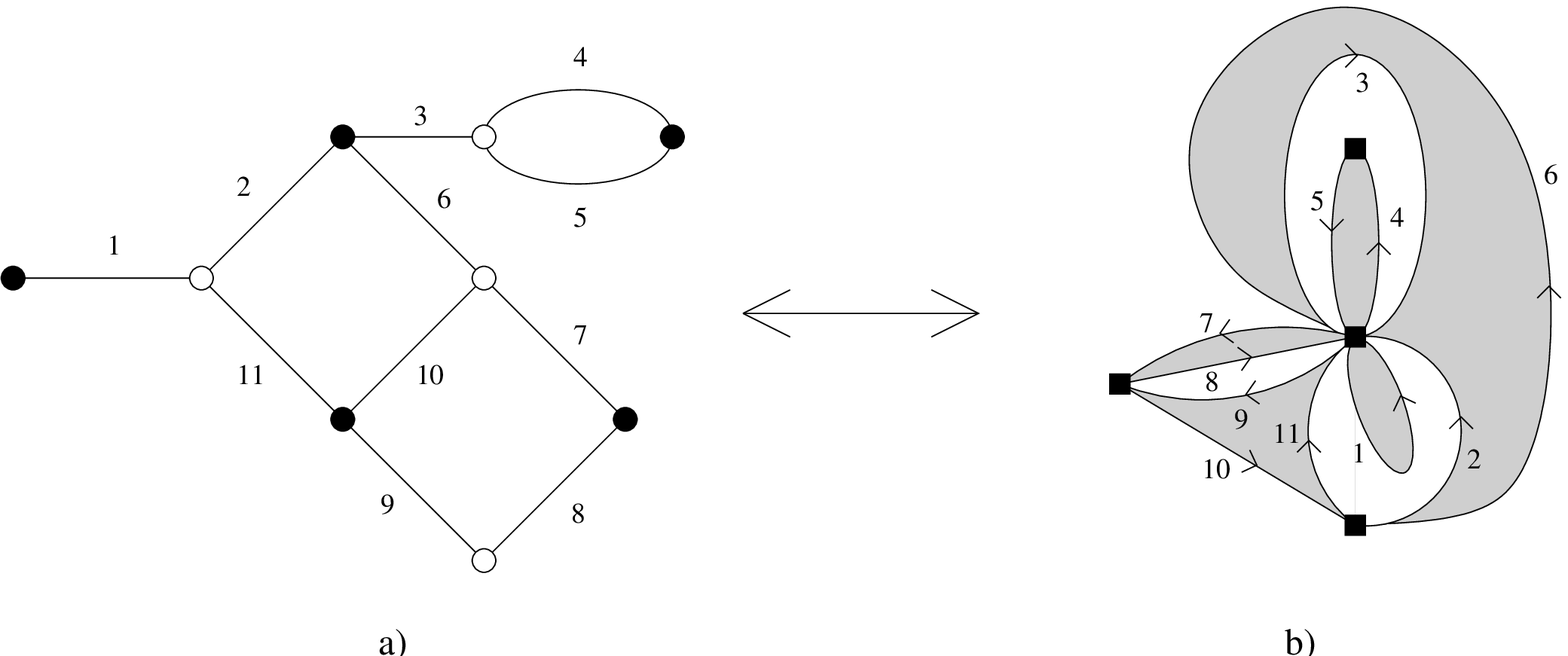}{12cm}

Also, for $\rho$ a permutation of the set of edges,
define $\Pi_\rho$ to be the partition of the {\it set} of edges
into the orbits (cycles) of $\rho$. For two such partitions
$\Pi$ and $\Pi'$, we say that $\Pi\le\Pi'$ iff every block of $\Pi$
is included in a block of $\Pi'$; and define $\Pi\vee\Pi'$ to be the
smallest common majorant. Also, call $\#\Pi$
the number of blocks of $\Pi$.
Finally, if $n$ is the number of edges,
define $C_\rho\vdash n$ to be the partition of the {\it integer} $n$
corresponding to $\Pi_\rho$, i.e.\ the lengths of the cycles of
$\rho$.

Then the results of \BC\ (Theorems 2.12, 4.2) can be reformulated
as follows:
\eqn\thBC{
F(A,B;s)=\sum_{n=1}^\infty 
 { s^{-n}\over n!}
\sum_{\scriptstyle \sigma,\tau\in\goth{S}_n\atop 
\scriptstyle \#\Pi_\sigma+\#\Pi_\tau+\#\Pi_{\sigma\tau}-n
=2\#(\Pi_\sigma\vee\Pi_\tau)
}
\gamma\left(\sigma\tau,\Pi_{\sigma} \vee
\Pi_{\tau}\right)\ 
\tr_{C_{\sigma}}A\ \,\tr_{C_{\tau}} B \ .}
The coefficient $\gamma(\rho,\Pi)$ is only defined explicitly
if $\Pi=\Pi_\rho$, in which case $\gamma(\rho,\Pi_\rho)=W_{C_\rho}$
where $W_\alpha$ is given by Eq.~\Wresu{b}.
In general, we have the following expression for $\gamma(\rho,\Pi)$:
\eqn\thBCb{
\gamma(\rho,\Pi)=
\sum_{\scriptstyle\Pi'\ge\Pi_\rho,  
\#(\Pi\vee\Pi')=1
\atop \scriptstyle\#\Pi_\rho-\#\Pi'=\#\Pi-1}
\prod_{\scriptstyle{\rm blocks\ }i\atop \scriptstyle{\rm of\ }\Pi'} W_{C_{\rho_i}}
}
where the $\rho_i$ are the restrictions of $\rho$ to each
block $i$ of the partition $\Pi'$. This expression looks somewhat
complicated but it is very easy to interpret
once we have related this formalism to the diagrammatic expansion
of the previous section.

In order to proceed with the equivalence, we first 
associate to the pair of permutations $\sigma$ and $\tau$ a map
according to the construction above. Since $\sigma$ and $\tau$
are permutations of $\{ 1,\ldots,n \}$, this produces a map
with {\it labelled edges}\/ from $1$ to $n$.
The quantity $\#\Pi_\sigma+\#\Pi_\tau+\#\Pi_{\sigma\tau}-n$
is simply the Euler--Poincar\'e characteristic of the map and the
condition on the summation simply imposes {\it planarity}\/ of the map
(or more precisely, of each of its connected components).

Secondly, we unlabel the resulting map. That is,
since the summand of Eq.~\thBC\ does not actually depend on the labelling,
one can sum together maps which are only distinguished by the labelling
of edges; by definition of the symmetry factor of a map
(inverse of the
order of the automorphism group, i.e.\ here number of permutations that
commute with both $\sigma$ and $\tau$),
the only modification this produces is to replace
the $1/n!$ with the symmetry factor of the unlabelled map.
Thus, we rewrite Eq.~\thBC:
\eqn\thBCc{
F(A,B,s)=\sum_g 
{s^{-n} 
\over |{\rm Aut} \, g|}\ 
\gamma\left(\sigma(g)\tau(g),\Pi_{\sigma(g)} \vee
\Pi_{\tau(g)}\right)\ 
\tr_{C_{\sigma(g)}}A\ \,\tr_{C_{\tau(g)}} B
}
where the summation is over inequivalent (possibly disconnected)
planar maps $g$, and $n$ is the
number of edges.

Thirdly, we replace these
maps $g$ with their dual maps $\hat{g}$. Note that this
must be performed separately for each connected component. 
The correspondence is again one-to-one. An example is given on Fig.~2.
The new maps have bicolored faces; equivalently, one can
orient their edges so that white (resp.\ black) vertices 
correspond to clockwise  (resp.\ counterclockwise) faces.

The resulting maps $\hat{g}$ are very similar to the Feynman diagrams
of sec.~5.1; however, they still lack the ``dotted lines'' 
which link together the various connected components.
This is where Eq.~\thBCb\ comes in. Indeed to each term in the
sum \thBCb\ we associate one particular set of dotted lines as follows:
since $\Pi'\ge \Pi_{\sigma(g)\tau(g)}$, and the cycles of $\sigma(g)\tau(g)$
are associated to faces of $g$, one can think of $\Pi'$ as ways of 
grouping together faces of $g$, that is vertices of $\hat{g}$:
these are precisely the dotted lines. Finally, the other conditions
in the summation of Eq.~\thBCb\ can be interpreted as follows:
$\#(\Pi' \vee \Pi_{\sigma(g)} \vee \Pi_{\tau(g)})=1$ ensures that the diagram 
including dotted lines is connected; and 
$\#\Pi_{\tau(g)\sigma(g)}-\#\Pi'=\#(\Pi_{\sigma(g)}\vee\Pi_{\tau(g)})-1$
ensures that it is ``minimally connected'', i.e.\ that it has a tree structure.

Finally, we compare the weights: the $W_{C_{\rho_i}}$ are associated to
groups of vertices of $\hat{g}$ linked together by dotted lines, just as in
the previous section; as to the $\tr_{C_{\sigma(g)}} A$ (resp.\ 
$\tr_{C_{\tau(g)}} B$), they associate to each white (resp.\ black)
vertex of $g$, that is each clockwise (resp.\ counterclockise) face of $\hat{g}$,
a weight of $\theta_p$ (resp.\ $\bar{\theta}_p$) where $p$ is the number
of edges surrounding it,
which is again what is required.

{\it Remark:} the special case 
$\Pi_{\sigma(g)}\vee \Pi_{\tau(g)}=\Pi_{\sigma(g)\tau(g)}$,
for which the associated weight is a single $W_{C_{\sigma(g)\tau(g)}}$,
occurs when $g$ is a disjoint union of bicolored trees, or
equivalently when all vertices
of $\hat{g}$ are linked together by dotted lines.

\subsec{Case of rectangular matrices}
\nind 
We sketch here how the diagrammatic expansion of sec.~5.1 can be generalized
to the case of the integral \rect\ over rectangular matrices. Introduce
complex rectangular $N_1\times N_2$ matrices $X$ and $Y$; then,
applying the same trick as above, we find
\def\Ad{A^\dagger}\def\Bd{B^\dagger}\def\Vd{V^\dagger}\def\Yd{Y^\dagger}
\eqnn\rectrick
$$\eqalignno{
&\int_{U(N_2)} \D U \int_{U(N_1)} \D V \exp {N\over s}
\Tr (AUB\Vd+V\Bd\Ud\Ad) \cr
&\ ={\rm const}\int \D X\D\Xd \D Y\D\Yd 
\e{-N'\Tr(X\Xd+Y\Yd)+ N_2^2 W(\Xd A\Ad Y;s)+N_1^2 W(\Bd\Yd XB;s)}
&\rectrick}$$
where $N=\min(N_1,N_2)$, $N'=\max(N_1,N_2)$.
The diagrammatic expansion is identical to that of sec.~5.1, except
that now there are two types of vertices, weighted by extra factors
$N_1/N'$ and $N_2/N'$ respectively. Due to the presence of $X$ and $\Yd$,
resp.\ $Y$ and $\Xd$,
in these vertices, we see that they must alternate and therefore
these diagrams have both faces {\it and}\/ vertices bicolored.
Up to a factor of 2 corresponding to the two possible colorings of the
vertices, this is the same as requiring diagrams to have bicolored even-sized
faces. Each face of size $2p$ receives a weight $\Tr (A\Ad)^p$ or
$\Tr (B\Bd)^p$ depending on its color (orientation).

Note that if $N_1=N_2=N$, the diagrams
are weighted identically as in sec.~5.1; this means that the integral
\newI\ for square matrices can be considered in the large $N$ limit
as the particular case of the integral \hciz\ for 
which all odd moments vanish (cf a similar observation in sec.~4.4).

\def\tc{\bar \phi}
\newsec{Summary of results, Tables}
\nind One may compute the first terms of the expansion 
$F(A,B;s)=\sum_{n=1}^\infty {1\over s^n} F_n(\s,\t)$.
It is sufficient to tabulate them for $\s_1=\t_1=0$ since
if we write $A=A'+\s_1 I$, $B=B'+\t_1 I$, with $A'$ and $B'$
traceless, $F(A,B;s)=F(A',B';s)+ \s_1\t_1/s\,.$
Or alternatively 
$$ F_n(\s_1,\t_1,\s_2,\t_2,\cdots)=\s_1\t_1 \Gd_{n1}+ F_n(0,0,
\s'_2,\t'_2,\cdots )$$
with $\s'_p=\sum_{q=0}^p {p \choose q}\s_q (-\s_1)^{p-q}$ and likewise for
$\t_p$. Up to order 8, we find:
\def\frac#1#2{{#1\over #2}}
\def\omit#1{{}}
\def\s{\theta}  
\def\t{\bar\theta}  
\overfullrule=0mm
%
%
%
{\baselineskip=0pt\lineskip=2pt
$$\eqalignno{
\scriptstyle
F_1&\scriptstyle =0\qquad
\scriptstyle  
F_2\scriptstyle =\frac{1}{2}\,\s_2\,\t_2\qquad 
\scriptstyle 
F_3\scriptstyle =\frac{1}{3}\,\s_3\,\t_3 \cr
\scriptstyle 
F_4&\scriptstyle =\frac{1 }{4}\left[\,{\s_2}^2\,\left( 3\,{\t_2}^2 - 
        2\,\t_4 \right)  + 
     \s_4\,\left( -2\,{\t_2}^2 + 
        \t_4 \right)\right]\cr
\scriptstyle 
F_5&\scriptstyle = \frac{1}{5}\,\left[\s_2\,\s_3\,\left( 20\,\t_2\,\t_3 - 
       5 \t_5 \right)  + 
    \s_5\,\left( -5\,\t_2\,\t_3 + 
         \t_5 \right)\right] \cr
\scriptstyle 
\omit{&\scriptstyle =(\s_2\,\s_3\,\left( 4\,\t_2\,\t_3 - \t_5 \right)  + 
    \frac{1 }{5}\s_5\,\left( -5\,\t_2\,\t_3 + 
         \t_5 \right)) \cr
\scriptstyle }
F_6&\scriptstyle =\frac{1 }{6} \big[
     {\s_2}^3\,\left( 27\,{\t_2}^3 - 
        16\,{\t_3}^2 - 
        30\,\t_2\,\t_4 + 
        7\,\t_6 \right)
+ {\s_3}^2\,\left( -16\,{\t_2}^3 + 
        6\,{\t_3}^2 + 
        15\,\t_2\,\t_4 - 
        3\,\t_6 \right) 
 \cr
\scriptstyle  &\scriptstyle  \quad
+      3\,\s_2\,\s_4\,\left( -10\,{\t_2}^3 + 
        5\,{\t_3}^2 + 
        10\,\t_2\,\t_4 - 
        2\,\t_6 \right) 
+      \s_6\,\left( 7\,{\t_2}^3 - 
        3\,{\t_3}^2 - 
        6\,\t_2\,\t_4 + 
        \t_6 \right) 
\big]
 \cr
\scriptstyle     
F_7&\scriptstyle =
   {\s_2}^2\,\s_3\,\left( 66\,{\t_2}^2\,\t_3
        - 21\,\t_2\,\t_5  
       -20\,\t_3\,\t_4 + 4 
          \t_7   \right)
\cr
\scriptstyle   &\scriptstyle \quad 
-     \s_3\,\s_4\,\left( 20\,{\t_2}^2\,
        \t_3 - 
       5\,\t_3\,\t_4 - 
       6\,\t_2\,\t_5 + 
       \t_7 \right)  
\cr
\scriptstyle   &\scriptstyle \quad 
-     \s_2\,\s_5\,\left( 21\,{\t_2}^2\,
        \t_3 - 
       6\,\t_3\,\t_4 - 
       6\,\t_2\,\t_5 + 
       \t_7 \right) 
\cr
\scriptstyle   &\scriptstyle \quad 
+\frac{1}{7}\, \s_7\,\left( 28\,{\t_2}^2\,
   \t_3 -   7\,\t_3\,\t_4 -  7\,\t_2\,\t_5 +  \t_7 \right) 
 \cr
\scriptstyle 
F_8&\scriptstyle =\frac{1}{8}\big[
     3\,{\s_2}^4\,\left( 117\,{\t_2}^4 
         -192\,  \t_2\,{\t_3}^2            - 180\,{\t_2}^2\,\t_4 
+  25\,{\t_4}^2  
+ 56\,\t_3\,\t_5    +  56\, \t_2\,\t_6   - 10\,\t_8 \right) 
 \cr
\scriptstyle  &\scriptstyle \quad - 
     4\,\s_2\, {\s_3}^2\,
         \left( 144\,{\t_2}^4 - 
           176\,\t_2\,{\t_3}^2 - 
           200\,{\t_2}^2\,\t_4 + 
           25\,{\t_4}^2 + 
           48\,\t_3\,\t_5 + 
           56\,\t_2\,\t_6 - 
           9\,\t_8 \right)
\cr
\scriptstyle   &\scriptstyle \quad 
- 4\,{\s_2}^2\,\s_4\,
      \left( 135\,{\t_2}^4 - 
        195\,{\t_2}^2\,\t_4 + 
        25\,{\t_4}^2 + 
        54\,\t_3\,\t_5 + 
        \t_2\,
         \left( -200\,{\t_3}^2 + 
           56\,\t_6 \right)  - 9\,\t_8 \ \right)   
\cr
\scriptstyle   &\scriptstyle \quad + {\s_4}^2\,\left( 75\,{\t_2}^4 - 
        100\,\t_2\,{\t_3}^2 - 
        100\,{\t_2}^2\,\t_4 + 
        10\,{\t_4}^2 + 
        24\,\t_3\,\t_5 + 
        28\,\t_2\,\t_6 - 
        4\,\t_8 \right)  
 \cr
\scriptstyle   &\scriptstyle \quad
+    \s_3\,\s_5\, \left( 168\,{\t_2}^4 
-      192\,\t_2\,{\t_3}^2 
-     216\,{\t_2}^2\,\t_4 +      24\,{\t_4}^2 
+      48\,\t_3\,\t_5 
+      56\,\t_2\,\t_6
-      8\,\t_8
\right) 
 \cr
\scriptstyle   &\scriptstyle \quad
+       \s_2\,  \s_6\,\left( 168\,{\t_2}^4 - 
           224\,\t_2\,{\t_3}^2 - 
           224\,{\t_2}^2\,\t_4 + 
           28\,{\t_4}^2 + 
           56\,\t_3\,\t_5 + 
           56\,\t_2\,\t_6 - 
           8\,\t_8 \right)  
 \cr
\scriptstyle   &\scriptstyle \quad\qquad  
 + \s_8\,\left(
       -  30\,{\t_2}^4     
+   36\,\t_2\,{\t_3}^2 
+     36\,{\t_2}^2\,\t_4 
-      4\,{\t_4}^2   
- 8\,\,\t_3\,\t_5 - 8\,\t_2\,\t_6 +\t_8 
\right) 
\big]\cr
}$$}

\omit{$$F_9=(105\,\s_4\,\s_5\,{\t_2}^3\,\t_3 - 
    \frac{55\,\s_9\,{\t_2}^3\,\t_3}{3} - 
    10\,\s_4\,\s_5\,{\t_3}^3 + 
    \frac{5\,\s_9\,{\t_3}^3}{3} - 
    60\,\s_4\,\s_5\,\t_2\,\t_3\,
     \t_4 + 10\,\s_9\,\t_2\,
     \t_3\,\t_4 - 
    33\,\s_4\,\s_5\,{\t_2}^2\,\t_5 + 
    5\,\s_9\,{\t_2}^2\,\t_5 + 
    6\,\s_4\,\s_5\,\t_4\,\t_5 - 
    \s_9\,\t_4\,\t_5 + 
    7\,\s_4\,\s_5\,\t_3\,\t_6 - 
    \s_9\,\t_3\,\t_6 + 
    8\,\s_4\,\s_5\,\t_2\,\t_7 - 
    \s_9\,\t_2\,\t_7 + 
    \frac{{\s_2}^3\,\s_3\,\left( 3780\,{\t_2}^3\,
          \t_3 - 416\,{\t_3}^3 - 
         1323\,{\t_2}^2\,\t_5 + 
         315\,\t_4\,\t_5 + 
         336\,\t_3\,\t_6 - 
         360\,\t_2\,
          \left( 7\,\t_3\,\t_4 - 
            \t_7 \right)  - 55\,\t_(9) \
\right) }{3} - \frac{{\s_3}^3\,
       \left( 416\,{\t_2}^3\,\t_3 - 
         32\,{\t_3}^3 - 
         240\,\t_2\,\t_3\,
          \t_4 - 
         144\,{\t_2}^2\,\t_5 + 
         30\,\t_4\,\t_5 + 
         28\,\t_3\,\t_6 + 
         36\,\t_2\,\t_7 - 
         5\,\t_(9) \right) }{3} + 
    \frac{\s_3\,\s_6\,\left( 336\,{\t_2}^3\,
          \t_3 - 28\,{\t_3}^3 - 
         189\,\t_2\,\t_3\,
          \t_4 - 
         105\,{\t_2}^2\,\t_5 + 
         21\,\t_4\,\t_5 + 
         21\,\t_3\,\t_6 + 
         24\,\t_2\,\t_7 - 
         3\,\t_(9) \right) }{3} - 
    \s_4\,\s_5\,\t_(9) + 
    \frac{\s_9\,\t_(9)}{9} + 
    {\s_2}^2\,\s_5\,\left( -441\,{\t_2}^3\,
        \t_3 + 48\,{\t_3}^3 + 
       144\,{\t_2}^2\,\t_5 - 
       33\,\t_4\,\t_5 - 
       35\,\t_3\,\t_6 + 
       6\,\t_2\,
        \left( 47\,\t_3\,\t_4 - 
          6\,\t_7 \right)  + 5\,\t_(9) \
\right)  + \s_2\,\left( \s_7\,
        \left( 120\,{\t_2}^3\,\t_3 - 
          12\,{\t_3}^3 - 
          36\,{\t_2}^2\,\t_5 + 
          8\,\t_4\,\t_5 + 
          8\,\t_3\,\t_6 + 
          8\,\t_2\,
           \left( -9\,\t_3\,\t_4 + 
             \t_7 \right)  - \t_(9) \
\right)  + \s_3\,\s_4\,\left( -840\,{\t_2}^3\,
           \t_3 + 80\,{\t_3}^3 + 
          282\,{\t_2}^2\,\t_5 - 
          60\,\t_4\,\t_5 - 
          63\,\t_3\,\t_6 + 
          \t_2\,
           \left( 515\,\t_3\,\t_4 - 
             72\,\t_7 \right)  + 10\,\t_(9) \
\right)  \right) )$$
$$F_{10}=(\frac{-441\,{\s_5}^2\,{\t_2}^5}{10} - 
    84\,\s_4\,\s_6\,{\t_2}^5 + 
    \frac{143\,\s_{10}\,{\t_2}^5}{10} + 
    99\,{\s_5}^2\,{\t_2}^2\,{\t_3}^2 + 
    210\,\s_4\,\s_6\,{\t_2}^2\,{\t_3}^2 - 
    33\,\s_{10}\,{\t_2}^2\,{\t_3}^2 + 
    72\,{\s_5}^2\,{\t_2}^3\,\t_4 + 
    140\,\s_4\,\s_6\,{\t_2}^3\,\t_4 - 
    22\,\s_{10}\,{\t_2}^3\,\t_4 - 
    18\,{\s_5}^2\,{\t_3}^2\,\t_4 - 
    35\,\s_4\,\s_6\,{\t_3}^2\,\t_4 + 
    \frac{11\,\s_{10}\,{\t_3}^2\,\t_4}{2} - 
    18\,{\s_5}^2\,\t_2\,{\t_4}^2 - 
    35\,\s_4\,\s_6\,\t_2\,{\t_4}^2 + 
    \frac{11\,\s_{10}\,\t_2\,{\t_4}^2}{2} - 
    36\,{\s_5}^2\,\t_2\,\t_3\,
     \t_5 - 77\,\s_4\,\s_6\,\t_2\,
     \t_3\,\t_5 + 
    11\,\s_{10}\,\t_2\,\t_3\,
     \t_5 + \frac{3\,{\s_5}^2\,{\t_5}^2}
     {2} + \frac{7\,\s_4\,\s_6\,{\t_5}^2}{2} - 
    \frac{\s_{10}\,{\t_5}^2}{2} - 
    21\,{\s_5}^2\,{\t_2}^2\,\t_6 - 
    42\,\s_4\,\s_6\,{\t_2}^2\,\t_6 + 
    \frac{11\,\s_{10}\,{\t_2}^2\,\t_6}{2} + 
    \frac{7\,{\s_5}^2\,\t_4\,\t_6}{2} + 
    7\,\s_4\,\s_6\,\t_4\,\t_6 - 
    \s_{10}\,\t_4\,\t_6 + 
    4\,{\s_5}^2\,\t_3\,\t_7 + 
    8\,\s_4\,\s_6\,\t_3\,\t_7 - 
    \s_{10}\,\t_3\,\t_7 + 
    \frac{9\,{\s_5}^2\,\t_2\,\t_8}{2} + 
    9\,\s_4\,\s_6\,\t_2\,\t_8 - 
    \s_{10}\,\t_2\,\t_8 + 
    {\s_2}^3\,\s_4\,\left( -1080\,{\t_2}^5 + 
       2025\,{\t_2}^3\,\t_4 - 
       620\,{\t_3}^2\,\t_4 + 
       72\,{\t_5}^2 + 
       42\,{\t_2}^2\,
        \left( 75\,{\t_3}^2 - 
          16\,\t_6 \right)  + 
       140\,\t_4\,\t_6 + 
       156\,\t_3\,\t_7 - 
       3\,\t_2\,
        \left( 200\,{\t_4}^2 + 
          437\,\t_3\,\t_5 - 
          55\,\t_8 \right)  - 22\,\t_(10) \
\right)  - \frac{{\s_5}^2\,\t_(10)}{2} - 
    \s_4\,\s_6\,\t_(10) + 
    \frac{\s_{10}\,\t_(10)}{10} - 
    \s_3\,\s_7\,\left( 96\,{\t_2}^5 - 
       156\,{\t_2}^3\,\t_4 + 
       36\,{\t_3}^2\,\t_4 - 
       4\,{\t_5}^2 - 
       8\,\t_4\,\t_6 + 
       {\t_2}^2\,
        \left( -216\,{\t_3}^2 + 
          44\,\t_6 \right)  - 
       8\,\t_3\,\t_7 + 
       \t_2\,
        \left( 40\,{\t_4}^2 + 
          80\,\t_3\,\t_5 - 
          9\,\t_8 \right)  + \t_(10) \
\right)  + \frac{{\s_3}^2\,\s_4\,
       \left( 720\,{\t_2}^5 - 
         1240\,{\t_2}^3\,\t_4 + 
         280\,{\t_3}^2\,\t_4 - 
         36\,{\t_5}^2 - 
         56\,{\t_2}^2\,
          \left( 30\,{\t_3}^2 - 
            7\,\t_6 \right)  - 
         70\,\t_4\,\t_6 - 
         72\,\t_3\,\t_7 + 
         \t_2\,
          \left( 325\,{\t_4}^2 + 
            672\,\t_3\,\t_5 - 
            90\,\t_8 \right)  + 
         11\,\t_(10) \right) }{2} + 
    {\s_2}^5\,\left( \frac{2754\,{\t_2}^5}{5} - 
       1080\,{\t_2}^3\,\t_4 + 
       360\,{\t_3}^2\,\t_4 - 
       \frac{441\,{\t_5}^2}{10} - 
       54\,{\t_2}^2\,
        \left( 32\,{\t_3}^2 - 
          7\,\t_6 \right)  - 
       84\,\t_4\,\t_6 - 
       96\,\t_3\,\t_7 + 
       \frac{9\,\t_2\,
          \left( 75\,{\t_4}^2 + 
            168\,\t_3\,\t_5 - 
            22\,\t_8 \right) }{2} + 
       \frac{143\,\t_(10)}{10} \right)  + 
    {\s_2}^2\,\left( {\s_3}^2\,
        \left( -1728\,{\t_2}^5 + 
          3150\,{\t_2}^3\,\t_4 - 
          840\,{\t_3}^2\,\t_4 + 
          99\,{\t_5}^2 + 
          2\,{\t_2}^2\,
           \left( 2158\,{\t_3}^2 - 
             511\,\t_6 \right)  + 
          210\,\t_4\,\t_6 + 
          216\,\t_3\,\t_7 - 
          \frac{5\,\t_2\,
             \left( 365\,{\t_4}^2 + 
               720\,\t_3\,\t_5 - 
               99\,\t_8 \right) }{2} - 
          33\,\t_(10) \right)  + 
       \frac{\s_6\,\left( 756\,{\t_2}^5 - 
            1344\,{\t_2}^3\,\t_4 + 
            392\,{\t_3}^2\,\t_4 - 
            42\,{\t_5}^2 - 
            84\,\t_4\,\t_6 + 
            {\t_2}^2\,
             \left( -2044\,{\t_3}^2 + 
               406\,\t_6 \right)  - 
            88\,\t_3\,\t_7 + 
            \t_2\,
             \left( 385\,{\t_4}^2 + 
               798\,\t_3\,\t_5 - 
               90\,\t_8 \right)  + 
            11\,\t_(10) \right) }{2} \right)  + 
    \frac{\s_2\,\left( \s_8\,\left( -198\,{\t_2}^5 + 
            330\,{\t_2}^3\,\t_4 - 
            90\,{\t_3}^2\,\t_4 + 
            9\,{\t_5}^2 + 
            45\,{\t_2}^2\,
             \left( 11\,{\t_3}^2 - 
               2\,\t_6 \right)  + 
            18\,\t_4\,\t_6 + 
            18\,\t_3\,\t_7 - 
            18\,\t_2\,
             \left( 5\,{\t_4}^2 + 
               10\,\t_3\,\t_5 - 
               \t_8 \right)  - 2\,\t_(10) \
\right)  + {\s_4}^2\,\left( 675\,{\t_2}^5 - 
            1200\,{\t_2}^3\,\t_4 + 
            325\,{\t_3}^2\,\t_4 - 
            36\,{\t_5}^2 - 
            70\,\t_4\,\t_6 + 
            {\t_2}^2\,
             \left( -1825\,{\t_3}^2 + 
               385\,\t_6 \right)  - 
            80\,\t_3\,\t_7 + 
            5\,\t_2\,
             \left( 65\,{\t_4}^2 + 
               144\,\t_3\,\t_5 - 
               18\,\t_8 \right)  + 
            11\,\t_(10) \right)  + 
         2\,\s_3\,\s_5\,\left( 756\,{\t_2}^5 - 
            1311\,{\t_2}^3\,\t_4 + 
            336\,{\t_3}^2\,\t_4 - 
            36\,{\t_5}^2 - 
            77\,\t_4\,\t_6 + 
            {\t_2}^2\,
             \left( -1800\,{\t_3}^2 + 
               399\,\t_6 \right)  - 
            80\,\t_3\,\t_7 + 
            6\,\t_2\,
             \left( 60\,{\t_4}^2 + 
               118\,\t_3\,\t_5 - 
               15\,\t_8 \right)  + 
            11\,\t_(10) \right)  \right) }{2}) $$} 
 
\omit{
\vfill\eject
In terms of planar connected cumulants
$$\eqalign{F_1&\scriptstyle =0  \cr
\scriptstyle  
F_2&\scriptstyle =\frac{1}{2}\, f(2)\,f'(2) \cr
\scriptstyle 
F_3&\scriptstyle =\frac{1}{3}\, f(3)\,f'(3) \cr
\scriptstyle 
F_4&\scriptstyle =\frac{1}{4}\, \left( f(4)\,\,{f'}(4) -{f(2)}^2\,{\,{f'}(2)}^2
\right) \cr
\scriptstyle 
F_5&\scriptstyle =\frac{1}{5}\,f(5)\,\,{f'}(5) - f(2)\,f(3)\,\,{f'}(2)\,\,{f'}(3) \
\cr
\scriptstyle 
F_6&\scriptstyle =\frac{1}{6}\,\Big(     f(6)\,\,{f'}(6)+
{f(2)}^3\,\left( 2\,{\,{f'}(2)}^3 - 
        {\,{f'}(3)}^2 \right)  - 
     3\,f(2)\,f(4)\,\left( {\,{f'}(3)}^2 + 
        2\,\,{f'}(2)\,\,{f'}(4) \right)  - 
     {f(3)}^2\,\left( {\,{f'}(2)}^3 + 
        3\,{\,{f'}(3)}^2 + 
        3\,\,{f'}(2)\,\,{f'}(4) \right)  
\Big)\cr
\scriptstyle 
&\scriptstyle = \frac{f(6)\,{f'}(6)}{6}
 - 
  \left( \frac{{f(3)}^2}{2} + f(2)\,f(4) \right) \,
   \left( \frac{{{f'}(3)}^2}{2} + 
     {f'}(2)\,{f'}(4) \right)
 - 
  \left( \frac{{f(2)}^3}{3} + \frac{{f(3)}^2}{2} \right) \,
   \left( \frac{{{f'}(2)}^3}{3} + 
     \frac{{{f'}(3)}^2}{2} \right) 
+ \Big(\frac{2}{3}\,{f(2)}^3\Big)\,
        \Big(\frac{2}{3}\,{{f'}(2)}^3\Big) \cr
\scriptstyle 
F_7&\scriptstyle =
    \frac{f(7)\,\,{f'}(7)}{7}
+
{f(2)}^2\,f(3)\,\,{f'}(3)\,
     \left( 3\,{\,{f'}(2)}^2 - \,{f'}(4) \right)  - 
    f(2)\,f(5)\,\left( \,{f'}(3)\,\,{f'}(4) + 
       \,{f'}(2)\,\,{f'}(5) \right)  - 
    f(3)\,f(4)\,\left( {\,{f'}(2)}^2\,\,{f'}(3) + 
       2\,\,{f'}(3)\,\,{f'}(4) + 
       \,{f'}(2)\,\,{f'}(5) \right)  \cr
\scriptstyle 
&\scriptstyle =   \frac{f(7)\,{f'}(7)}{7}
 - 
  \big( f(3)\,f(4) + f(2)\,f(5) \big) \,
   \big( {f'}(3)\,{f'}(4) + 
     {f'}(2)\,{f'}(5) \big)
 - 
  f(3)\,\left( {f(2)}^2 + f(4) \right) \,{f'}(3)\,
   \left( {{f'}(2)}^2 + {f'}(4) \right) 
+
4\,{f(2)}^2\,f(3)\,{{f'}(2)}^2\,{f'}(3)
 \cr
\scriptstyle  
F_8&\scriptstyle =
\frac{1}{8}\Big[  f(8)\,\,{f'}(8)\,
-8\,f(3)\,f(5)\,\left(
\,{f'}(2)\,{\,{f'}(3)}^2 + 
      {\,{f'}(2)}^2\,\,{f'}(4)  
     + \,{\,{f'}(4)}^2 +2\,{f'}(3)\,\,{f'}(5) +\,{f'}(2)\,\,{f'}(6) \right) - 
      {f(2)}^4\,\left( 5\,{\,{f'}(2)}^4 - 
         8\,\,{f'}(2)\,{\,{f'}(3)}^2 + 
         {\,{f'}(4)}^2 \right)  \cr
\scriptstyle 
&\scriptstyle \quad + 
      4\,{f(2)}^2\,f(4)\,\left( 6\,\,{f'}(2)\,
          {\,{f'}(3)}^2 + 
         7\,{\,{f'}(2)}^2\,\,{f'}(4) - 
         {\,{f'}(4)}^2 - 
         2\,\,{f'}(3)\,\,{f'}(5) \right)  
-       {f(4)}^2\,\left( {\,{f'}(2)}^4 + 
         4\,{\,{f'}(2)}^2\,\,{f'}(4) + 
         6\,{\,{f'}(4)}^2 + 
         8\,\,{f'}(3)\,\,{f'}(5) + 
         4\,\,{f'}(2)\,
          \left( 2\,{\,{f'}(3)}^2 + \,{f'}(6) \
\right)  \right) \cr
\scriptstyle 
&\scriptstyle \quad + 8\,f(2)\,{f(3)}^2\,
          \left( {\,{f'}(2)}^4 + 
            5\,\,{f'}(2)\,{\,{f'}(3)}^2 + 
            3\,{\,{f'}(2)}^2\,\,{f'}(4) - 
            {\,{f'}(4)}^2 - 
            \,{f'}(3)\,\,{f'}(5) \right)  
-     4\,f(2)\,      f(6)\,\left( {\,{f'}(4)}^2 + 
            2\,\,{f'}(3)\,\,{f'}(5) + 
            2\,\,{f'}(2)\,\,{f'}(6) \right)  
\Big] \cr
\scriptstyle 
&\scriptstyle = \frac{f(8)\,{f'}(8)}{8}
  - 
  \left( \frac{{f(4)}^2}{2} + f(3)\,f(5) + f(2)\,f(6) \right) \,
   \left( \frac{{{f'}(4)}^2}{2} + 
     {f'}(3)\,{f'}(5) + 
     {f'}(2)\,{f'}(6) \right)\cr
\scriptstyle  
&\scriptstyle \quad
   - 
  \left( f(2)\,{f(3)}^2 + {f(2)}^2\,f(4) + \frac{{f(4)}^2}{2} + 
     f(3)\,f(5) \right) \,\left( {f'}(2)\,
      {{f'}(3)}^2 + 
     {{f'}(2)}^2\,{f'}(4) + 
     \frac{{{f'}(4)}^2}{2} + 
     {f'}(3)\,{f'}(5) \right)\cr
\scriptstyle 
&\scriptstyle \quad
 - \left( \frac{{f(2)}^4}{4} + f(2)\,{f(3)}^2 + 
       \frac{{f(4)}^2}{2} \right) \,
     \left( \frac{{{f'}(2)}^4}{4} + 
       {f'}(2)\,{{f'}(3)}^2 + 
       \frac{{{f'}(4)}^2}{2} \right)  
\scriptstyle  &\scriptstyle \quad 
+   \frac{f(2)\,\left( 8\,{f(3)}^2 + 9\,f(2)\,f(4) \right) \,
     {f'}(2)\,\left( 8\,{{f'}(3)}^2 + 
       9\,{f'}(2)\,{f'}(4) \right) }{18}\cr
\scriptstyle 
&\scriptstyle \quad +   \frac{f(2)\,\left( 45\,{f(2)}^3 + 124\,{f(3)}^2 \right) \,
     {f'}(2)\,\left( 45\,{{f'}(2)}^3 + 
       124\,{{f'}(3)}^2 \right) }{4464}
-\frac{63\,{f(2)}^4\,{{f'}(2)}^4}{62}  
}$$
\vfill\eject
$$\eqalign{F_{9}&\scriptstyle =    \frac{f(9)\,{f'}(9)}{9}
 -  \left( f(4)\,f(5) + f(3)\,f(6) + f(2)\,f(7) \right) \,
   \left( {f'}(4)\,{f'}(5) + 
     {f'}(3)\,{f'}(6) + 
     {f'}(2)\,{f'}(7) \right) \cr
\scriptstyle 
&\scriptstyle  - \left( \frac{{f(3)}^3}{3} + 
     \left( {f(2)}^2 + f(4) \right) \,f(5) + 
     f(3)\,\left( 2\,f(2)\,f(4) + f(6) \right)  \right) \,
   \left( \frac{{{f'}(3)}^3}{3} + 
     \left( {{f'}(2)}^2 + {f'}(4) \right) \,
      {f'}(5) + {f'}(3)\,
      \left( 2\,{f'}(2)\,{f'}(4) + 
        {f'}(6) \right)  \right) \cr
\scriptstyle 
&\scriptstyle  - 
  \left(  {f(2)}^3\,f(3)   + \frac{2\,{f(3)}^3}{3} + 
     2\,f(2)\,f(3)\,f(4) + f(4)\,f(5) \right) \,
   \left(  {{f'}(2)}^3\,{f'}(3) \
  + \frac{2\,{{f'}(3)}^3}{3} +
     2\,{f'}(2)\,{f'}(3)\,
      {f'}(4) + {f'}(4)\,{f'}(5) \
\right)\cr
\scriptstyle 
&\scriptstyle  + 
  \frac{\left( \frac{4\,{f(3)}^3}{3} + 9\,f(2)\,f(3)\,f(4) + 
       5\,{f(2)}^2\,f(5) \right) \,
     \left( \frac{4\,{{f'}(3)}^3}{3} + 
       9\,{f'}(2)\,{f'}(3)\,
        {f'}(4) + 
       5\,{{f'}(2)}^2\,{f'}(5) \right) }{5} \cr
\scriptstyle 
&\scriptstyle + 
  \frac{f(3)\,\left( 25\,{f(2)}^3 + 9\,{f(3)}^2 + 32\,f(2)\,f(4) \right) \,
     {f'}(3)\,\left( 25\,{{f'}(2)}^3 + 
       9\,{{f'}(3)}^2 + 
       32\,{f'}(2)\,{f'}(4) \right) }{80}\cr
\scriptstyle 
&\scriptstyle  + 
  \frac{f(3)\,\left( 25\,{f(2)}^3 + 41\,{f(3)}^2 \right) \,
     {f'}(3)\,\left( 25\,{{f'}(2)}^3 + 
       41\,{{f'}(3)}^2 \right) }{1968} \cr
\scriptstyle 
&\scriptstyle  +\frac{-1984\,{f(2)}^3\,f(3)\,{{f'}(2)}^3\,
     {f'}(3)}{123}
\cr
\scriptstyle }$$
\vfill\eject
$$\eqalign{F_{10}&\scriptstyle =  \frac{f(10)\,{f'}(10)}{10}
-   \left( \frac{{f(5)}^2}{2} + f(4)\,f(6) + f(3)\,f(7) + f(2)\,f(8) \
\right) \,\left( \frac{{{f'}(5)}^2}{2} + 
     {f'}(4)\,{f'}(6) + 
     {f'}(3)\,{f'}(7) + 
     {f'}(2)\,{f'}(8) \right)  
\cr
\scriptstyle  \quad &\scriptstyle 
 - 
  \left( {f(3)}^2\,f(4) + f(2)\,{f(4)}^2 + \frac{{f(5)}^2}{2} + 
     {f(2)}^2\,f(6) + f(4)\,f(6) 
        + f(3)\,\left( 2\,f(2)\,f(5) + f(7) \right)  \right) \,
   \left( {{f'}(3)}^2\,{f'}(4) + 
     {f'}(2)\,{{f'}(4)}^2 + 
     \frac{{{f'}(5)}^2}{2} + 
     {{f'}(2)}^2\,{f'}(6) + 
     {f'}(4)\,{f'}(6) + 
     {f'}(3)\,\left( 2\,{f'}(2)\,
         {f'}(5) + {f'}(7) \right)  \right)  
\cr
\scriptstyle  \quad &\scriptstyle 
 -   \left( \frac{3\,{f(2)}^2\,{f(3)}^2}{2} + {f(2)}^3\,f(4) + 
     2\,{f(3)}^2\,f(4) + f(2)\,{f(4)}^2 + 2\,f(2)\,f(3)\,f(5) 
+      \frac{{f(5)}^2}{2} + f(4)\,f(6) \right) \,
   \left( \frac{3\,{{f'}(2)}^2\,{{f'}(3)}^2}
      {2} + {{f'}(2)}^3\,{f'}(4) + 
     2\,{{f'}(3)}^2\,{f'}(4) + 
     {f'}(2)\,{{f'}(4)}^2 + 
     2\,{f'}(2)\,{f'}(3)\,
      {f'}(5) + \frac{{{f'}(5)}^2}{2} + 
     {f'}(4)\,{f'}(6) \right) 
\cr
\scriptstyle  \quad &\scriptstyle 
 + 
  \frac{\left( 9\,{f(3)}^2\,f(4) + 10\,f(2)\,{f(4)}^2 + 
       20\,f(2)\,f(3)\,f(5) + 11\,{f(2)}^2\,f(6) \right) \,
     \left( 9\,{{f'}(3)}^2\,{f'}(4) + 
       10\,{f'}(2)\,{{f'}(4)}^2 + 
       20\,{f'}(2)\,{f'}(3)\,
        {f'}(5) + 
       11\,{{f'}(2)}^2\,{f'}(6) \right) }{22}
\cr
\scriptstyle  \quad &\scriptstyle 
 - 
  \left( \frac{{f(2)}^5}{5} + \frac{3\,{f(2)}^2\,{f(3)}^2}{2} + 
     {f(3)}^2\,f(4) + f(2)\,{f(4)}^2 + \frac{{f(5)}^2}{2} \right) \,
   \left( \frac{{{f'}(2)}^5}{5} + 
     \frac{3\,{{f'}(2)}^2\,{{f'}(3)}^2}{2} + 
     {{f'}(3)}^2\,{f'}(4) + 
     {f'}(2)\,{{f'}(4)}^2 + 
     \frac{{{f'}(5)}^2}{2} \right) 
\cr
\scriptstyle  \quad &\scriptstyle 
+ \frac{\left( 165\,{f(2)}^2\,{f(3)}^2 + 121\,{f(2)}^3\,f(4) + 
       119\,{f(3)}^2\,f(4) + 65\,f(2)\,{f(4)}^2 + 152\,f(2)\,f(3)\,f(5) \
\right) \,\left( 165\,{{f'}(2)}^2\,
        {{f'}(3)}^2 + 
       121\,{{f'}(2)}^3\,{f'}(4) + 
       119\,{{f'}(3)}^2\,{f'}(4) + 
       65\,{f'}(2)\,{{f'}(4)}^2 + 
       152\,{f'}(2)\,{f'}(3)\,
        {f'}(5) \right) }{1672}
\cr
\scriptstyle  \quad &\scriptstyle 
 + \frac{\left( 912\,{f(2)}^5 + 5765\,{f(2)}^2\,{f(3)}^2 + 
       225\,{f(2)}^3\,f(4) + 1735\,{f(3)}^2\,f(4) + 2985\,f(2)\,{f(4)}^2 \
\right) \,\left( 912\,{{f'}(2)}^5 + 
       5765\,{{f'}(2)}^2\,{{f'}(3)}^2 + 
       225\,{{f'}(2)}^3\,{f'}(4) + 
       1735\,{{f'}(3)}^2\,{f'}(4) + 
       2985\,{f'}(2)\,{{f'}(4)}^2 \right) }{2268600} 
\cr
\scriptstyle  \quad &\scriptstyle 
 + \frac{\left( 300\,{f(2)}^5 + 3491\,{f(2)}^2\,{f(3)}^2 + 
       726\,{f(2)}^3\,f(4) + 4192\,{f(3)}^2\,f(4) \right) \,
     \left( 300\,{{f'}(2)}^5 + 
       3491\,{{f'}(2)}^2\,{{f'}(3)}^2 + 
       726\,{{f'}(2)}^3\,{f'}(4) + 
       4192\,{{f'}(3)}^2\,{f'}(4) \right) }{2502624}
\cr
\scriptstyle  \quad &\scriptstyle 
 - 
  \frac{3\,{f(2)}^2\,\left( 124\,{f(2)}^3 + 17135\,{f(3)}^2 + 
       13966\,f(2)\,f(4) \right) \,{{f'}(2)}^2\,
     \left( 124\,{{f'}(2)}^3 + 
       17135\,{{f'}(3)}^2 + 
       13966\,{f'}(2)\,{f'}(4) \right) }{29272736}
\cr
\scriptstyle  \quad &\scriptstyle 
 -   \frac{3\,{f(2)}^2\,\left( 15638\,{f(2)}^3 + 66833\,{f(3)}^2 \right) \,
     {{f'}(2)}^2\,
     \left( 15638\,{{f'}(2)}^3 + 
       66833\,{{f'}(3)}^2 \right) }{466694839}
+\frac{872496\,{f(2)}^5\,{{f'}(2)}^5}{334165}
\cr
}$$
}

One may slightly simplify this expansion by making the following
asymmetric change of variables: we introduce instead of the moments
$\t_q$, the {\it free cumulants}\/ $\tc_q$ which are defined by 
\refs{\BIPZ,\IZ,\Spe}
\eqn\freec
{\tc_q=-\sum_{\alpha_1,\ldots,\alpha_q\ge 0\atop\sum_i i\alpha_i=q} 
{(q+\sum_i \alpha_i-2)!\over (q-1)!} 
\prod_i {(-\t_i)^{\alpha_i}\over\alpha_i!}\ .}
Indeed, if one now expands $F$ as 
$$F=\sum_{q\ge1} 
{\der F\over\der\theta_q}\Big|_{\theta=0} \theta_q 
+{1\over2!}\sum_{q,r\ge 1} 
{\der^2 F\over\der\theta_q\der\theta_r}\Big|_{\theta=0}
\theta_q\theta_r 
+\cdots
$$
then the first 3 derivatives were computed exactly in \PZJ\ using
the formalism of dispersionless Toda hierarchy, and turn
out to be simply expressible in terms of the $\tc_q$. In particular,
${\der F\over\der\theta_q}\big|_{\theta=0}={1\over q}\tc_q$.
The same expansion to order 8 now takes the form:\goodbreak
\def\frac#1#2{{#1\over #2}}
\def\s{\theta}
\def\tc{\bar \phi}
\def\({\left(}\def\){\right)}
%
%
{\baselineskip=0pt\lineskip=2pt
$$\eqalignno{\scriptstyle
F_2&\scriptstyle=\frac{1}{2}\s_2\tc_{2}\qquad
\scriptstyle
F_3\scriptstyle=\frac{1}{3}\s_3\tc_{3}\qquad
\scriptstyle
F_4\scriptstyle=\frac{1}{4}\s_4\tc_4 -\frac{1}{2}{\s_2^2} \left({\tc_4}
+\frac{1}{2}{\tc_2^2}\right) \cr
\scriptstyle
F_5&\scriptstyle=\frac{1}{5}\s_5\tc_5 -\s_3\s_2(\tc_5+ \tc_3\tc_2)\cr
\scriptstyle
F_6&\scriptstyle=\frac{1}{6}\s_6\tc_6 -\s_4\s_2
\left(\tc_6+\tc_4\tc_2+\frac{1}{2}{\tc_3^2}\right)-\frac{1}{2}{\s_3^2}
\left(\tc_6+\tc_4\tc_2+\tc_3^2 +\frac{1}{3}{\tc_2^3}\right)
+\frac{1}{6}{\s_2^3}\(7\tc_6+12 \tc_4\tc_2+5\tc_3^2+2\tc_2^3\)
\cr
\scriptstyle 
F_7&\scriptstyle=\frac{1}{7}\s_7\tc_7 -\s_5\s_2\(\tc_7+\tc_5\tc_2+\tc_4\tc_3\)
-\s_4\s_3\(\tc_7+\tc_5\tc_2+2\tc_4\tc_3+\tc_3\tc_2^2\)+\s_3\s_2^2\(
4\tc_7+7\tc_5\tc_2+8\tc_4\tc_3+5\tc_3\tc_2^2\)\cr
\scriptstyle
F_8&\scriptstyle=\frac{1}{8}\s_8\tc_8
-\s_6\s_2\(\tc_8+\tc_6\tc_2+\tc_5\tc_3+\frac{1}{2}{\tc_4^2}\)
-\s_5\s_3\(\tc_8+\tc_6\tc_2+2\tc_5\tc_3+{\tc_4^2}+\tc_4\tc_2^2+\tc_3^2\tc_2\)
\cr
\scriptstyle
&\scriptstyle \qquad -\frac{1}{2}{\s_4^2}\(\tc_8+\tc_6\tc_2+2\tc_5\tc_3+
\frac{3}{2}{\tc_4^2}+{\tc_4\tc_2^2}+2\tc_3^2\tc_2+\frac{1}{4}{\tc_2^4}\)\cr
\scriptstyle
&\scriptstyle\qquad
+\frac{1}{2}{\s_4\s_2^2}\(9\tc_8+16\tc_6\tc_2+18\tc_5\tc_3 +11\tc_4^2
+11\tc_4\tc_2^2+14\tc_3^2\tc_2+\tc_2^4\)\cr
\scriptstyle
&\scriptstyle\qquad+\s_3^2\s_2\(
9\tc_8+16\tc_6\tc_2+24\tc_5\tc_3 +11\tc_4^2
+16\tc_4\tc_2^2+20\tc_3^2\tc_2+2\tc_2^4
\)\cr
\scriptstyle
&\scriptstyle\qquad -\frac{3}{8}{\s_2^4}
\(10\tc_8+24\tc_6\tc_2+24\tc_5\tc_3 +15\tc_4^2
+24\tc_4\tc_2^2+24\tc_3^2\tc_2+3\tc_2^4 \)
\cr}
$$}
\omit{
$$\eqalignno{\scriptstyle
F_9&\scriptstyle=\frac{1}{9}\s_9\tc_9
-\s_7 \s_2\(
\tc_4 \tc_5 + \tc_3 \tc_6 + \tc_2 \tc_7 + \tc_9\)
-\s_6 \s_3 \(
\frac{\tc_3^3}{3} + \tc_2^2 \tc_5 + 2 \tc_4 \tc_5 
  +  2 \tc_3\ \tc_6 +2 \tc_2 \tc_3 \tc_4 +2\tc_2 \tc_7 + \tc_9\)
\cr
\scriptstyle &\scriptstyle \qquad
-\s_5\s_4\(
\tc_2^3 \tc_3 + \tc_3^3 + \tc_2^2 \tc_5 + 3 \tc_4 \tc_5 + 
    2 \tc_3 \tc_6 + 4 \tc_2 \tc_3 \tc_4 +\tc_2 \tc_7 + \tc_9\)
\cr
\scriptstyle &\scriptstyle \qquad
+{\s_5\s_2^2\over 2}
\(4 \tc_2^3 \tc_3 + 6 \tc_3^3 + 12 \tc_2^2 \tc_5 + 
    24 \tc_4 \tc_5 + 20 \tc_3 \tc_6 + 
   30\tc_2 \tc_3 \tc_4 + 18\tc_2 \tc_7 + 10 \tc_9\)
\cr
\scriptstyle &\scriptstyle \qquad
+\s_4\s_3\s_2
\(13 \tc_2^3\ \tc_3 + 11 \tc_3^3 + 18 \tc_2^2 \tc_5 + 
    30 \tc_4 \tc_5 + 27 \tc_3 \tc_6 + 
    +53 \tc_2 \tc_3 \tc_4 + 18\tc_2 \tc_7 + 10 \tc_9\)
\cr
\scriptstyle &\scriptstyle \qquad
+\frac{\s_3^3}{6}
\(16\tc_2^3\tc_3 + 16\tc_3^3 + 24\tc_2^2\tc_5 + 
    34\tc_3\tc_6 + 60\tc_2\tc_3\tc_4 + 18\tc_2\tc_7 + 
    30\tc_4\tc_5 +10 \tc_9\)
\cr
\scriptstyle &\scriptstyle \qquad
-\frac{\s_3\s_2^3}{6}\(
210\tc_2^3\tc_3 + 136\tc_3^3 + 306\tc_2^2\tc_5 + 
    360\tc_4\tc_5 + 318\tc_3\tc_6 + 
    738\tc_2 \tc_3\tc_4 + 270\tc_2\tc_7 + 110\tc_9\)
\cr
\scriptstyle 
F_{10}&\scriptstyle=
\frac{1}{10}\s_{10}\tc_{10}
-\s_8\s_2
\(\frac{ \tc_{5}^2}{2} + \tc_{4}\tc_{6} + \tc_{3}\tc_{7} + \tc_{2}\tc_{8} + 
    \tc_{10}\)
\cr
\scriptstyle &\scriptstyle \  \tc_{2}
-\s_7\s_3
\(\tc_{3}^2\tc_{4} + \tc_{5}^2 + \tc_{2}^2\tc_{6} + 
    2\tc_{4}\tc_{6} + 2\tc_{3}\tc_{7} + 
    \tc_{2}\tc_{4}^2 +2 \tc_{2}\tc_{3}\tc_{5} + \tc_{2}\tc_{8} + 
    \tc_{10}\)
\cr
\scriptstyle &\scriptstyle \ 
-\s_6\s_4
\(\tc_{2}^3\tc_{4} + 3\tc_{3}^2\tc_{4} + \frac{3\tc_{5}^2}{2} +
     {3 \tc_{2}^2\tc_{3}^2\over 2}  + \tc_{2}^2\tc_{6} + 3\tc_{4}\tc_{6} + 
      2\tc_{3}\tc_{7} + 
     2 \tc_{2}\tc_{4}^2 +4 \tc_{2}\tc_{3}\tc_{5} + \tc_{2}\tc_{8} + 
      \tc_{10}\)
\cr
\scriptstyle &\scriptstyle \ 
-{\s_5^2\over 2}
\(\frac{\tc_{2}^5}{5} + 3\tc_{2}^2\tc_{3}^2 + \tc_{2}^3\tc_{4} + 
      4\tc_{3}^2\tc_{4} + 3\tc_{2}\tc_{4}^2 + 4\tc_{2}\tc_{3}\tc_{5} + 
      2\tc_{5}^2 + \tc_{2}^2\tc_{6} + 3\tc_{4}\tc_{6} + 
      2\tc_{3}\tc_{7} + \tc_{2}\tc_{8} + \tc_{10}\)
\cr
\scriptstyle &\scriptstyle \ 
+{\s_6\s_2^2\over 2}
\(4\tc_{2}^3\tc_{4} + 19\tc_{3}^2\tc_{4} + 13\tc_{5}^2 + 
    26\tc_{4}\tc_{6} + 6\tc_{2}^2\tc_{3}^2 + 13\tc_{2}^2\tc_{6} + 
    22\tc_{3}\tc_{7} + 
    16\tc_{2}\tc_{4}^2 + 32\tc_{2}\tc_{3}\tc_{5} + 20\tc_{2}\tc_{8}
         + 11\tc_{10}\)
\cr
\scriptstyle &\scriptstyle \ 
\s_5\s_3\s_2
\(\tc_{2}^5 + 14\tc_{2}^3\tc_{4} + 40\tc_{3}^2\tc_{4} + 19\tc_{5}^2 + 
    33\tc_{4}\tc_{6} + 27\tc_2^2\tc_{3}^2 + 20\tc_2^2\tc_{6} + 
    30\tc_{3}\tc_{7} + 
33  \tc_{2}\tc_{4}^2 + 58 \tc_{2}\tc_{3}\tc_{5} + 20 \tc_{2}\tc_{8}
 + 11\tc_{10}\)
\cr
\scriptstyle &\scriptstyle \!\!\!\!\!\! 
+{\s_4^2 \s_2\over 2}
\(2\tc_{2}^5 + 20\tc_{2}^3\tc_{4} + 50\tc_{3}^2\tc_{4} + 19\tc_{5}^2 + 
    40\tc_{4}\tc_{6} +40\tc_{2}^2\tc_{3}^2 + 20\tc_{2}^2 \tc_{6} + 
    30\tc_{3}\tc_{7} + 
   40\tc_{2}\tc_{4}^2 + 70\tc_{2}\tc_{3}\tc_{5} + 20\tc_{2}\tc_{8}
 + 11\tc_{10}\)
\cr
\scriptstyle &\scriptstyle \!\!\!\!\!\! 
+{\s_4 \s_3^2\over 2}
\(2\tc_{2}^5 + 22\tc_{2}^3\tc_{4} + 61\tc_{3}^2\tc_{4} + 19\tc_{5}^2 + 
    40\tc_{4}\tc_{6} +44 \tc_{2}^2\tc_{3}^2 + 27 \tc_{2}^2\tc_{6} + 
    38\tc_{3}\tc_{7} + 
  40\tc_{2}\tc_{4}^2 + 78\tc_{2}\tc_{3}\tc_{5} + 20\tc_{2}\tc_{8} 
        + 11\tc_{10}\)
\cr
\scriptstyle &\scriptstyle \!\!\!\!\!\! 
-{\s_4\s_2^3}
\(4\tc_{2}^5 + 47\tc_{2}^3\tc_{4} + 
   85 \tc_{2}^2\tc_{3}^2 + 62 \tc_{2}^2\tc_{6} + 
  90 \tc_{2}\tc_{4}^2 + 159 \tc_{2}\tc_{3}\tc_{5} + 55 \tc_{2}\tc_{8} + 
    98\tc_{3}^2\tc_{4} + 38\tc_{5}^2 + 80\tc_{4}\tc_{6} + 
          64\tc_{3}\tc_{7} + 22\tc_{10}\)
\cr
\scriptstyle &\scriptstyle \!\!\!\!\!\! 
-{\s_3^2\s_2^2\over 2}
\(18\tc_{2}^5 + 184\tc_{2}^3\tc_{4} + 
   338 \tc_{2}^2\tc_{3}^2 + 214 \tc_{2}^2\tc_{6} + 
295\tc_{2}\tc_{4}^2 + 576\tc_{2}\tc_{3}\tc_{5} + 165\tc_{2}\tc_{8} + 
    366\tc_{3}^2\tc_{4} + 132\tc_{5}^2 + 240\tc_{4}\tc_{6} + 
          228\tc_{3}\tc_{7} + 66\tc_{10}\)
\cr
\scriptstyle &\scriptstyle \!\!\!\!\!\! 
+{\s_2^5\over 10}
\(54\tc_{2}^5 + 855\tc_{2}^2\tc_{3}^2 + 540\tc_{2}^3\tc_{4} + 
    795\tc_{3}^2\tc_{4} + 810\tc_{2}\tc_{4}^2 + 1380\tc_{2}\tc_{3}\tc_{5} + 
    274\tc_{5}^2 + 615\tc_{2}^2\tc_{6} + 590\tc_{4}\tc_{6} + 
    470\tc_{3}\tc_{7} + 440\tc_{2}\tc_{8} + 143\tc_{10}\)
}$$}

One notes the intriguing feature that in either basis, for all known
terms (up to $p=10$), $p F_p$ has only integer coefficients. 
In other words, the quantity $s\partial F/\partial s $ has
only integer coefficients. The rationale behind this fact and 
its reinterpretation in terms of the counting of ``rooted'' objects
has remained elusive to us. The integral \hciz\ still has a few mysteries\dots

In this article, we have presented the standard lore 
on the integral \hciz\ as well as recent developments. 
The integral itself is  well understood by a variety of
methods (sect.~2) and admits interesting generalizations (sect.~3).
However the explicit expression, in terms of symmetric functions
of the eigenvalues, of the integral itself (not to mention 
the more complicated problem
of the associated correlation functions) is more subtle;
sect.~4 was devoted to the discussion of this problem
in the large $N$ limit.
These considerations, based on integrable hierarchies,
are complemented by other recent work based 
on combinatorial arguments, as discussed in sect.~5.  
It is our feeling that
the exact interrelations between these various approaches
require further study.
\vskip1cm
\centerline{\bf Acknowledgments}
{It is a pleasure to thank M. Bauer for his help in the reading of \DH.
One of the authors (J-BZ) wants to acknowledge the hospitality
of MSRI, Berkeley, where part of this work was performed.}

\listrefs
\bye